\documentclass{aa}  
\usepackage{graphicx}
\usepackage{txfonts}
\usepackage{aalongtable}
\usepackage[normalem]{ulem}
\usepackage{amstext}
\usepackage{color}

\begin{document}

\input{psfig}
%
\newbox\grsign \setbox\grsign=\hbox{$>$} \newdimen\grdimen \grdimen=\ht\grsign
\newbox\simlessbox \newbox\simgreatbox
\setbox\simgreatbox=\hbox{\raise.5ex\hbox{$>$}\llap
     {\lower.5ex\hbox{$\sim$}}}\ht1=\grdimen\dp1=0pt
\setbox\simlessbox=\hbox{\raise.5ex\hbox{$<$}\llap
     {\lower.5ex\hbox{$\sim$}}}\ht2=\grdimen\dp2=0pt
\def\simgreat{\mathrel{\copy\simgreatbox}}
\def\simless{\mathrel{\copy\simlessbox}}
\newbox\simppropto
\setbox\simppropto=\hbox{\raise.5ex\hbox{$\sim$}\llap
     {\lower.5ex\hbox{$\propto$}}}\ht2=\grdimen\dp2=0pt
\def\simpropto{\mathrel{\copy\simppropto}}
\makeatletter
\renewcommand{\table}[1][]{\@float{table}[!htp]}
\makeatother 
\title{High-resolution abundance analysis of red giants in  the
metal-poor bulge globular cluster HP~1
\thanks{Observations collected at the European Southern Observatory,
Paranal, Chile (ESO), under programs 93.D-0124A,
and 65.L-0340A}}


\author{
B. Barbuy\inst{1}
\and
E. Cantelli\inst{1}
\and
A. Vemado\inst{1}
\and
H. Ernandes\inst{1}
\and
S. Ortolani\inst{2,3}
\and
I. Saviane\inst{4}
\and
E. Bica\inst{5}
\and
D. Minniti\inst{6,7}
\and
B. Dias\inst{4}
\and
Y. Momany\inst{3}
\and
V. Hill\inst{8}
\and
M. Zoccali\inst{6,9}
\and
C. Siqueira-Mello\inst{1}
}
\offprints{B. Barbuy}

\institute{
Universidade de S\~ao Paulo, IAG, Rua do Mat\~ao 1226,
Cidade Universit\'aria, S\~ao Paulo 05508-900, Brazil;
e-mail: barbuy@astro.iag.usp.br
\and
Dipartimento di Fisica e Astronomia, Universit\`a di Padova, I-35122 Padova,
 Italy
\and
INAF-Osservatorio Astronomico di Padova, Vicolo dell'Osservatorio 5,
I-35122 Padova, Italy
\and
European Southern Observatory, Alonso de Cordova 3107, Santiago, Chile
\and
Universidade Federal do Rio Grande do Sul, Departamento de Astronomia,
CP 15051, Porto Alegre 91501-970, Brazil
\and
Millenium Institute of Astrophysics, Av. Vicu\~na Mackenna 4860,
Macul, Santiago, Chile
\and
Departamento de Ciencias Fisicas, Universidad Andres Bello, Republica 220, 
Santiago, Chile
\and
Laboratoire Lagrange (UMR7293), Universit\'e de Nice Sophia Antipolis,
 CNRS, Observatoire de la C\^ote d’Azur, CS 34229, F-06304 Nice cedex 4, France
\and
Pontificia Universidad Catolica de Chile, Instituto de Astrofisica,
Casilla 306, Santiago 22, Chile
}
 
   \date{Received; accepted }

 \abstract
   {The globular cluster HP~1 is projected at only  3\fdg33 from the Galactic center.
Together with its distance, this makes it one of the most central globular clusters in the Milky Way.
It has a blue horizontal branch (BHB) and a metallicity of 
[Fe/H]$\approx-1.0.$ This means that it probably is one of the 
oldest objects in the Galaxy.  
Abundance ratios can reveal the nucleosynthesis pattern of the first stars as well as the early chemical
enrichment and early formation of stellar populations.
 }
   {High-resolution spectra obtained for six
stars were analyzed to derive the abundances of the light elements
C, N, O, Na, and Al, the alpha-elements Mg, Si, Ca, and Ti, 
and the heavy elements Sr, Y , Zr, Ba, La, and Eu.}
   {High-resolution spectra of six red giants that are confirmed members of
  the bulge globular cluster HP~1 were obtained with the 8m VLT UT2-Kueyen 
 telescope with the UVES spectrograph in FLAMES-UVES
configuration.  The spectroscopic parameter derivation was based on the excitation
 and ionization equilibrium of \ion{Fe}{I} and \ion{Fe}{II}.}
   { We confirm a mean metallicity  of
[Fe/H] = $-1.06\pm 0.10$, by adding the two stars that were previously analyzed in HP~1.
The alpha-elements O and Mg are enhanced by about 
$+0.3\simless$~[O,Mg/Fe]~$\simless+0.5$~dex, 
Si is moderately enhanced with $+0.15\simless$~[Si/Fe]~$\simless+0.35$~dex,
 while Ca and Ti show lower values of 
  $-0.04\simless$~[Ca,Ti/Fe]~$\simless+0.28$~dex. 
  The r-element Eu is also enhanced with [Eu/Fe]$\approx+0.4$, which together
with O and Mg is indicative of early enrichment by type II supernovae.
Na and Al are low, but it is unclear if 
Na-O are anticorrelated. The heavy elements are moderately enhanced, with 
$-0.20<$~[La/Fe]~$<+0.43$~dex and $0.0<$~[Ba/Fe]~$<+0.75$~dex, 
which is compatible with r-process formation. The spread in Y, Zr, Ba, and La abundances,
 on the other hand, appears to be compatible with the spinstar scenario or 
 other additional mechanisms such as the weak r-process. }
   {  }
\titlerunning{Abundance analysis of 6 giants in HP~1}
\authorrunning{B. Barbuy et al.}
   \maketitle
%

\section{Introduction}

The globular cluster HP~1 is located at 3\fdg33 and 1.8 kpc from
the Galactic center. This is in the inner bulge volume, and the
cluster lies among the globular clusters
that are closest to the Galactic center. 
A metallicity of [Fe/H]~$\sim-1.0$ was deduced from its 
color-magnitude diagram (CMD) by Ortolani et al.
(1997, 2011).
High-resolution spectroscopy of two stars performed by Barbuy et al. (2006) 
resulted in [Fe/H]=$-1.0\pm$0.2, and low-resolution spectroscopy
of eight red giants by Dias et al. (2016) yielded a mean of
[Fe/H]=$-1.17\pm$0.07. 
Metallicities like this are at the lower
end of the metal-poor stellar population in the metallicity distribution
function (MDF) of bulge stars by 
Zoccali et al. (2008), Hill et al. (2011) and Rojas-Arriagada et al. (2014).  
A lower metallicity end at [Fe/H]$\sim$-1.0
 is due to the fast chemical enrichment 
in the Galactic bulge as
modeled for example by Cescutti et al. (2008, A\&A, 491, 401).
 There are traces of a very metal-poor population in the bulge,
such as those found by Garc\'{i}a-Perez et al. (2013) and
Howes et al. (2014, 2015). These stars are very interesting,
but most of them are located in the outer bulge and might more
probably be halo stars. The main fact is that the bulk of the bulge
stars show a lower end at [Fe/H]~$\sim-1.0$.
A metallicity of [Fe/H]~$\sim-1.0$
 could correspond to the population C  or D
as defined by Ness et al. (2013a) in an MDF of
a large sample of bulge stars.  For a
latitude b=-5$^{\circ}$ , Ness et al. found
a mean [Fe/H]=-0.66 for population C stars and identified it with the
thick-disk population; their population D has a mean [Fe/H]=-1.16
and was identified by them as a metal-weak thick-disk
population. It is not clear whether HP 1
 fits into these categories.

The following evidence shows that there may be a stellar
population peak at [Fe/H]~$\sim-1.0$ in the bulge:
a) The metallicity distribution of bulge  globular clusters
was shown to have two peaks at  
[Fe/H]~$\approx-0.5$ and [Fe/H]~$\approx-1.0$ 
(Bica et al. 2016), where a list of known bulge clusters
was selected. This had already been pointed out in
Barbuy et al. (2006, 2007, 2009) and Rossi et al. (2015).
b) Another piece of evidence that stars of this metallicity are very old
and are characteristic of the old Galactic bulge are the findings by
Walker \& Terndrup (1991),
who showed that RR Lyrae in the bulge show a metallicity peak at
 [Fe/H]~$\approx-1.0$ (see also Lee 1992).
  Based on the MACHO survey,
 Kunder \& Chaboyer (2008) determined a mean [Fe/H]=-1.25,
with a broad metallicity distribution.
 D\'ek\'any et al. (2013) used the VISTA Variables in
 the Via Lactea (VVV) survey 
and obtained a spheroidal and centrally concentrated distribution,
with a slight elongation in its center. Using OGLE-III data,
 Pietrukowicz et al. (2012)
 derived [Fe/H]=-1.02$\pm$0.18, with a barred distribution
toward the central parts of the Galaxy. From OGLE-IV
Pietrukowicz et al.  (2015) obtained a mean [Fe/H]=-1.025$\pm$0.25 and
 a triaxial ellipsoid shape.
The outer bulge
studied with 10$^3$ VVV RR Lyrae by Gran et al. (2015)
 shows a centrally concentrated spheroidal distribution.  
c) More recently, Schultheis et al. (2015) found a peak of bulge
field stars at [Fe/H]~$\sim-1.0$ that was enhanced in alpha-elements.
d) Schiavon et al. (2016, in preparation) identified a sample of nitrogen-rich
stars that also show metallicities of [Fe/H]~$\sim-1.0$. 

 The triaxial ellipsoid shape as well as a
 cylindrical rotation as found in the bulge radial velocity
assay (BRAVA) by Kunder et al. (2012) 
are expected from the dynamical evolution of an initially 
small classical bulge and a bar formed later in the disk
(Saha et al. 2012). Therefore either a spheroidal or triaxial
shape would be consistent with an initially small classical bulge.

In summary, the moderately metal-poor globular clusters
in the inner Galactic bulge might be relics of an
early generation of long-lived stars formed in the proto-Galaxy. 
For this reason we have been pursuing the study of these clusters
based on spectroscopy (Barbuy et al. 2006, 2007, 2009, 2014), as well as
photometry and color-magnitude diagrams (CMDs) corrected for proper motion,
as can be found in Ortolani et al. (2011) and Rossi et al. (2015).

In the present work we carry out a detailed
analysis with  high spectral resolution of six stars of the globular cluster HP~1. This cluster has a 
blue horizontal branch (BHB) combined with a metallicity of [Fe/H]$\sim-1.0$, 
which indicates that it is very old.
 Ortolani et al. (2011)
derived the age by plotting HP~1 in the diagram of Fig. 17 by
 Dotter et al. (2010). From this, they computed an age difference of
about 1 Gyr for HP 1 compared to their sample of halo clusters
with $\sim$12.7 Gyr. This means that HP 1 is about 13.7 Gyr old
and appears to be one of the oldest globular clusters in the Galaxy.

The cluster HP~1 is located at J2000 
 $\alpha=17^{\rm h}31^{\rm m}05.2^{\rm s}$, $\delta=-29^{\rm o}58'54\arcsec$, 
 with Galactic coordinates l~$=-2\fdg58$, b~$=+2\fdg12$.

The globular cluster HP~1 was discovered at the Observatoire
de Haute Provence by Dufay et al. (1954).
It was first studied through CMDs by Ortolani et al. (1997)
by means of V, I colors. 
 Davidge (2000) studied individual stars in the J, H, K, and CO filters
and estimated [Fe/H]=$-$1.6 for HP~1. 
Ortolani et al. (2011) used J, H, and K with the multiconjugate
adaptive optics demonstrator (MAD) at the VLT, applying
 a proper motion decontamination procedure,
making use of the time difference between the NTT observations from 1994
and the VLT/MAD observations in 2008. This allowed producing decontaminated
CMDs and computing the orbit of HP~1 in the Galaxy.
The CMD proper motion cleaning greatly optimizes the selection of 
member stars. 
 Ortolani et al. (2011)  showed that HP~1 remains confined within the bulge
and/or bar.
Minniti (1995) employed medium-resolution infrared spectroscopy and measured
indices in six stars of HP~1. They obtained a metallicity of [Fe/H]=$-$0.56
and a radial velocity of 60 km.s$^{-1}$.
 Stephens et al. (2004) used medium-resolution infrared spectra of
six stars in HP~1 and derived a metallicity of [Fe/H]=$-$1.30.
 Two stars of HP~1 were observed at high spectral resolution
 with UVES and were analyzed spectroscopically
 (Barbuy et al. 2006), program 65.L-0340 (PI: D. Minniti).
 To provide the most accurate information on the abundance
pattern and kinematics of the metal-poor bulge globular clusters, 
we analyze a more significant number of stars in HP~1 to better identify the  characteristics of the
 probably oldest stellar population in  the Galaxy.

In this work we present a detailed abundance analysis using data from
the FLAMES-UVES spectrograph at the VLT  with a resolution R~$\sim$~45,000 
and a signal-to-noise ratio S/N~$>200$ for all sample stars.
The MARCS model atmospheres are employed (Gustafsson et al. 2008).

   The observations are described  in Sect. 2.
  The  effective temperature of the  photometric stellar parameters
and the gravity  
are   derived in  Sect. 3.  Spectroscopic parameters are
derived in  Sect.  4 and   abundance ratios are   computed in  Sect.
5. A discussion is presented in Sect. 6 and conclusions are drawn in Sect. 7.

\section{Observations} 

\begin{figure}
\centering
\psfig{file=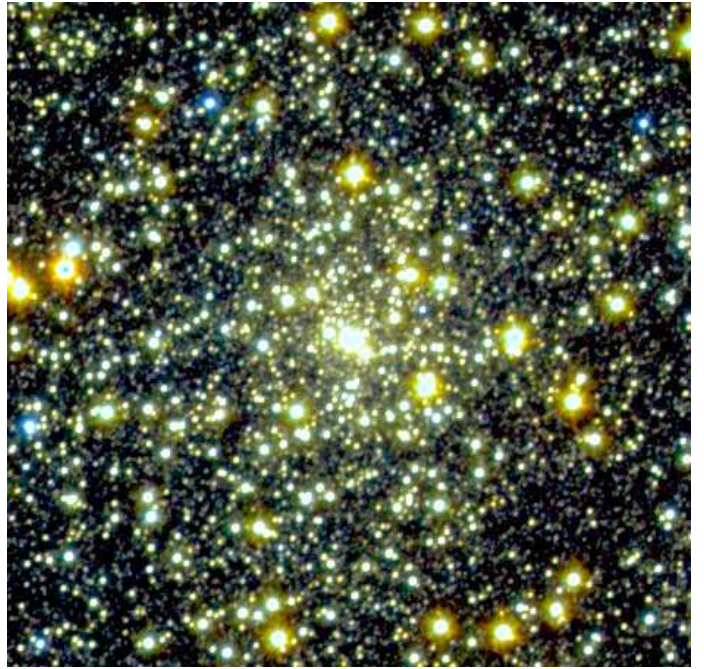,angle=0.,width=9.0 cm}
\caption{HP~1 JHK$_{s}$-combined colour image from the VVV survey.
The image has size of 2$\times$2 arcmin$^{2}$. North is at 45$^{\circ}$ anticlockwise.}
\label{imageJHK} 
\end{figure}

\begin{figure}
\centering
\psfig{file=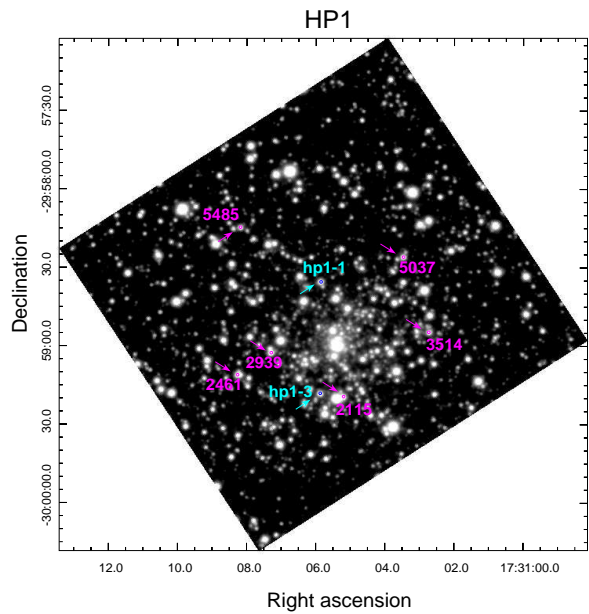,angle=0.,width=9.0 cm}
\caption{z image from the VVV survey, indicating
the location of the sample stars.
}
\label{image} 
\end{figure}

The sample member stars of HP~1
 were selected from data corrected for the proper motion that
were reported in Ortolani et al. (2011).
Figure \ref{imageJHK} shows a JHK$_{s}$-combined image
of HP~1 from the Vista Variables in the Via Lactea VVV survey
 (Saito et al. 2012)\footnote{http://horus.roe.ac.uk/vsa/}.
 The location of the six sample
stars together with the two stars previously analyzed in Barbuy 
et al. (2006) are shown in Fig. \ref{image}, with the observed field
in a z-color image from the VVV survey.

The spectra of individual stars of HP~1
 were obtained at the VLT using the UVES spectrograph
(Dekker et al. 2000) in FLAMES-UVES mode. 
 The red arm (5800$-$6800 {\rm \AA}) has the
ESO CCD \# 20 chip, an MIT backside illuminated, with a size of 4096~x~2048 pixels, and a pixel
size  of 15~x~15~$\mu$m.
The blue arm (4800$-$5800 {\rm \AA}) uses the ESO Marlene EEV
CCD\#44 chip, backside illuminated, with a size of 4102~x~2048 pixels, 
and a pixel size of  15~x~15$\mu$m. 
The UVES standard setup 580 yields a resolution R~$\sim$~45,000 for a slit width of 1
arcsec. 
  The pixel scale is $0.0147$~{\rm \AA}/pix, 
with $\sim7.5$ pixels  per resolution element at 6000 {\rm \AA}.
 The data were reduced using the 
UVES pipeline within the ESO/Reflex software (Ballester et al. 2000;
Modigliani et al. 2004). 
The log of the 2014 observations is given in  Table \ref{logobs}. 
 The   spectra   were  flat fielded,  optimally extracted,   and
wavelength calibrated    with the  FLAMES-UVES pipeline.
Spectra  extracted  from different frames were  then  co-added, taking into 
account the radial velocities reported in Table \ref{vr}. 
The present UVES observations centered on 5800 {\rm \AA} yield
a spectral coverage of $4800<\lambda<6800$ {\rm \AA},
with a gap at 5708$-$5825 {\rm \AA}.


 We measured the radial velocities of each run using 
the IRAF FXCOR cross-correlation method, as
reported in Table \ref{vr}.
The mean  errors reported from the IRAF routine are 
1.52 km~s$^{\rm {-1}}$. These values are clearly higher
than jitter variations, which are estimated to be around 0.1 km~s$^{\rm {-1}}$
(e.g., Hekker et al. 2008).
 
The present mean heliocentric radial velocity 
v$^{\rm hel}_{\rm r}=+40.0\pm0.5$~km~s$^{\rm {-1}}$ 
agrees very well with the value
of  $+45.8$~km~s$^{\rm {-1}}$ derived from two stars in Barbuy et al. (2006).
These are very low radial velocities, which combined with the low
proper motions (Ortolani et al. (2011) support the possibility
of a confinement in the bulge.
                                 
\begin{figure}
\centering
\psfig{file=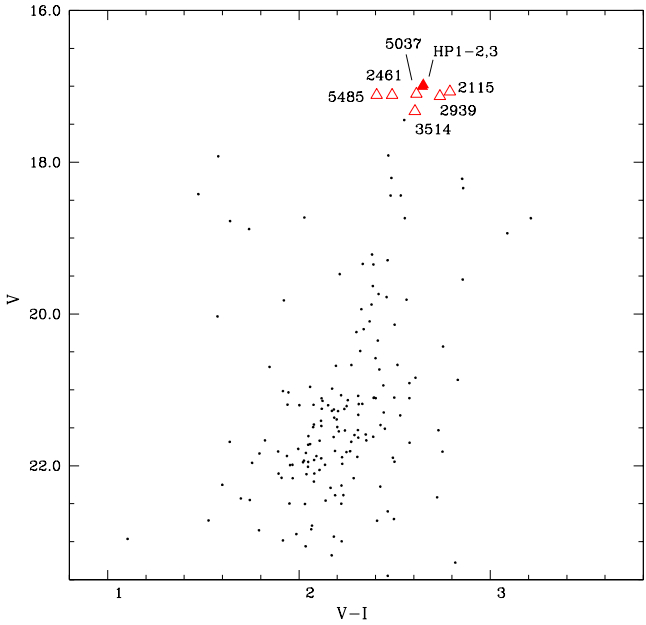,angle=0.,width=9.0 cm}
\caption{V, I CMD of HP~1 based on data from Ortolani et al. (1997),
with the V, V$-$I of the sample stars indicated.
}
\label{cmd} 
\end{figure}

 Figure \ref{cmd} shows the 
V, I CMD of HP~1 and the location of the program stars on the
red giant branch.

\begin{table*}
\caption[1]{Log of the spectroscopic observations: dates, Julian dates,
exposure times, airmass, seeing, and run number.}
\begin{flushleft}
\begin{tabular}{lllllllccccccc}
\noalign{\smallskip}
\hline
\noalign{\smallskip}
\hline
\noalign{\smallskip}
Date & UT & HJD & exp   & Airmass  
& Seeing & run  & \\ 
\noalign{\smallskip}
 &  &  & (s) &   & ($''$) & &  \\
\noalign{\smallskip}
\noalign{\smallskip}
\hline
\noalign{\vskip 0.2cm}
 07.06.14 & 04:33:15 & 2456815.69535 & 3000  & 1.007-1.013 & 0.72-0.84 &1& \\
 24.06.14 & 01:55:50 & 2456832.58620 & 2700  & 1.083-1.026 & 0.47-0.73  &2& \\
 20.07.14 & 05:13:40 & 2456858.72267 & 2700  & 1.329-1.591 & 0.52-0.53  &3& \\
 30.07.14 & 01:26:09 & 2456868.56407 & 2700  & 1.004-1.023 & 0.83-0.75  &4& \\
 14.08.14 & 01:00:33 & 2456883.54517 & 2750  & 1.010-1.047 & 0.71-0.69  &5& \\
 14.08.14 & 01:58:18 & 2456883.54517 & 2750  & 1.048-1.126 & 0.64-0.72  &6& \\
 14.08.14 & 02:45:54 & 2456883.61832 & 2750  & 1.129-1.264 & 0.77-1.00  &7& \\
\noalign{\smallskip}
\hline
\end{tabular}
\end{flushleft}
\label{logobs}
\end{table*}

\begin{table}
\caption[2]{Radial velocities of the UVES sample stars in each of the seven
exposure runs, corresponding heliocentric radial velocities, and mean
heliocentric radial velocity.}
\small
\begin{flushleft}
\begin{tabular}{l@{}c@{}c@{}c@{}c@{}c@{}c@{}c@{}}
\noalign{\smallskip}
\hline
\noalign{\smallskip}
\hline
\noalign{\smallskip}
Target &
 \phantom{-}${\rm v_r^{obs}}$ & \phantom{-}\phantom{-}${\rm v_r^{hel.}}$ & & \phantom{-}Target & \phantom{-}${\rm v_r^{obs}}$ & 
\phantom{-}\phantom{-}${\rm v_r^{hel.}}$ \\
\noalign{\smallskip}
&${\rm km~s^{-1}}$ &${\rm km~s^{-1}}$ & & &${\rm km~s^{-1}}$ &${\rm km~s^{-1}}$ \\
\noalign{\smallskip}
\noalign{\smallskip}
\hline
\noalign{\vskip 0.2cm}
2115 1 & \phantom{-}+37.024 & \phantom{-}+41.004 & & \phantom{-}2461 1 & \phantom{-}+36.430 & \phantom{-}+40.41 & \\ 
2115 2 & \phantom{-}+45.313 &  \phantom{-}+41.143 & & \phantom{-}2461 2 &\phantom{-}+44.920 & \phantom{-}+40.75 &  \\ 
2115 3 & \phantom{-}+57.650 & \phantom{-}+41.140 & & \phantom{-}2461 3 & \phantom{-}+57.695 & \phantom{-}+41.185 & \\ 
2115 4 & \phantom{-}+60.847 & \phantom{-}+40.797 & & \phantom{-}2461 4 & \phantom{-}+61.178 & \phantom{-}+41.128 &  \\ 
2115 5 & \phantom{-}+65.612 & \phantom{-}+40.812 & & \phantom{-}2461 5 & \phantom{-}+65.902 & \phantom{-}+41.102 & \\ 
2115 6 & \phantom{-}+65.699 & \phantom{-}+40.799 & & \phantom{-}2461 6 & \phantom{-}+65.995 & \phantom{-}+41.095 & \\ 
2115 7 & \phantom{-}+65.819 & \phantom{-}+40.839 & & \phantom{-}2461 7 & \phantom{-}+66.137 & \phantom{-}+41.157 & \\ 
\hbox{Mean} & --- & \phantom{-}+41.006 & &
 \hbox{Mean} & --- & \phantom{-}+41.048 & \\
\noalign{\smallskip}
\hline
\noalign{\vskip 0.2cm}
2939 1 & \phantom{-}+41.756 &  \phantom{-}+45.736 & & \phantom{-}3514 1 & \phantom{-}+31.695 & \phantom{-}+35.675 & \\
2939 2 & \phantom{-}+50.289 &  \phantom{-}+46.119 & & \phantom{-}3514 2 & \phantom{-}+39.904 & \phantom{-}+35.734 & \\
2939 3 & \phantom{-}+62.512 & \phantom{-}+46.002 & & \phantom{-}3514 3 & \phantom{-}+52.524 & \phantom{-}+36.014 & \\
2939 4 & \phantom{-}+66.057 & \phantom{-}+46.007 & & \phantom{-}3514 4 & \phantom{-}+55.573 & \phantom{-}+35.523 &  \\
2939 5 & \phantom{-}+70.718 & \phantom{-}+45.918 & & \phantom{-}3514 5 & \phantom{-}+59.992 & \phantom{-}+35.192 & \\
2939 6 & \phantom{-}+70.859 & \phantom{-}+45.959 & & \phantom{-}3514 6 & \phantom{-}+60.104 & \phantom{-}+35.204 & \\
2939 7 & \phantom{-}+70.974 & \phantom{-}+45.994 & & \phantom{-}3514 7 & \phantom{-}+60.168 & \phantom{-}+35.188 & \\
\hbox{Mean} & & \phantom{-}+46.035 & &\hbox{Mean} & --- & \phantom{-}+35.577 & \\
\noalign{\smallskip}
\hline
\noalign{\vskip 0.2cm}
5037 1 & \phantom{-}+37.355 &  \phantom{-}+41.335 & & \phantom{-}5485 1 & \phantom{-}+30.851 & \phantom{-}+34.831 & \\
5037 2 & \phantom{-}+45.627 &  \phantom{-}+41.457 & & \phantom{-}5485 2 & \phantom{-}+38.234 & \phantom{-}+34.064 & \\
5037 3 & \phantom{-}+58.229 & \phantom{-}+41.719 & & \phantom{-}5485 3 & \phantom{-}+50.745 & \phantom{-}+34.235 & \\
5037 4 & \phantom{-}+61.994 & \phantom{-}+41.994 & & \phantom{-}5485 4 & \phantom{-}+54.665 & \phantom{-}+34.615 &  \\
5037 5 & \phantom{-}+66.700 & \phantom{-}+41.900 & & \phantom{-}5485 5 & \phantom{-}+59.565 & \phantom{-}+34.765 & \\
5037 6 & \phantom{-}+66.684 & \phantom{-}+41.784 & & \phantom{-}5485 6 & \phantom{-}+59.782 & \phantom{-}+34.882 & \\
5037 7 & \phantom{-}+66.978 & \phantom{-}+41.998 & & \phantom{-}5485 7 & \phantom{-}+59.832 & \phantom{-}+34.852 & \\
\hbox{Mean} & & \phantom{-}+41.807 & &\hbox{Mean} & --- & \phantom{-}+34.679 & \\
\noalign{\smallskip}
\hline
\end{tabular}
\end{flushleft}
\label{vr}
\end{table}

\begin{table*}
\caption[1]{Identifications, coordinates, V, I magnitudes from Ortolani et al. (1997), and
$JHK_{s}$ magnitudes from the 2MASS and VVV surveys.
The two stars analyzed in Barbuy et al. (2006) are also listed.
$^{a}$: identification of star 2461 in VVV may be a blend of stars.}
\small
\begin{flushleft}
\tabcolsep 0.15cm
\begin{tabular}{ccccccccccccccccccc}
\noalign{\smallskip}
\hline
\noalign{\smallskip}
\hline
\noalign{\smallskip}
{\rm  ID no.}& 2MASS ID & $\alpha_{2000}$ & $\delta_{2000}$ & $V$ & $I$ & $J$ & $H$ & $K_{\rm s}$ &   $J_{\rm VVV}$ 
& {\rm H$_{\rm VVV}$} & {\rm K$_{\rm VVV}$} &  \cr
\noalign{\vskip 0.2cm}
\noalign{\hrule\vskip 0.2cm}
\noalign{\vskip 0.2cm}
 2115 & --- - --- & 17:31:05.420& $-$29:59:20.71 & 17.070 & 14.281 &--- & --- & --- & --- & --- &--- & \\
2461   & 17310822-2959108   & 17:31:08.22& $-$29:59:10.85 & 17.115 & 14.628 & --- & --- & --- &12.066$^a$ &10.249$^a$ &10.959$^a$ & \\
 2939 & 17310729-2959021  & 17:31:07.31& $-$29:59:02.30 & 17.131 & 14.394 &11.901 &10.869 &10.595 &--- &--- &--- & \\
 3514 & 17310273-2958544 & 17:31:02.74& $-$29:58:54.74 & 17.327 & 14.721 &12.447 &11.574 &11.212 & --- &11.248 &11.378 & \\
 5037 & 17310347-2958259 &17:31:03.48& $-$29:58:26.04 & 17.101 & 14.487 &12.302 &11.303 &10.994 &12.287 &11.502 &11.115 & \\
 5485 & 17310817-2958147 & 17:31:08.19& $-$29:58:14.81 & 17.116 & 14.710 &12.761 &11.965 &11.364&12.737 &11.989 &11.729 & \\

\hline
 hp1-2 & 17310585-2958354 & 17:31:05.852&  $-$29:58:35.5 & 16.982 & 14.332 
&12.21 &11.268 &10.969  &12.1527 &11.4164 &11.0541 & \\
 hp1-3 & 17310587-2959180 & 17:31:05.872&  $-$29:59:18.1 & 17.003 & 14.358 
&12.167 &11.146 &10.828  &12.1220 &10.9702 &10.9119 & \\
 \hline
\end{tabular}
\end{flushleft}
\label{starmag}
\end{table*}


\begin{table*}
\caption[1]{
Photometric stellar parameters derived using the calibrations by 
Alonso et al. (1999) for $V-I$, $V-K$, $J-K$, bolometric corrections
computed by adopting the V, I derived temperature, bolometric magnitudes,
and corresponding gravity log $g$,
and final spectroscopic parameters. A reddening of  $E(B-V)=1.12$ is assumed.
For each star the last columns give
the spectroscopic parameters from UVES spectra (present work).} 
\small
\begin{flushleft}
\begin{tabular}{c@{}c@{}c@{}c@{}c@{}c@{}c@{}c@{}cc@{}c@{}c@{}c@{}c@{}c@{}c@{}c@{}c}
\noalign{\smallskip}
\hline
\noalign{\smallskip}
\hline
\noalign{\smallskip}
& \multicolumn{4}{c}{\hbox{}} Photometric\phantom{-} parameters & & & & \multicolumn{7}{c}{\hbox{}} Spectroscopic\phantom{-} parameters\\
\cline{2-9}  \cline{11-18}    \\
{\rm star} & \phantom{-}\phantom{-}\phantom{-}\phantom{-}T$_{\rm V-I}$ 
& \phantom{-}\phantom{-}\phantom{-}\phantom{-}T$_{\rm V-K}$ 
& \phantom{-}\phantom{-}\phantom{-}\phantom{-}T$_{\rm J-K}$
 &\phantom{-}\phantom{-}\phantom{-}\phantom{-}T$_{\rm V-K}$ 
 &  \phantom{-}\phantom{-}\phantom{-}\phantom{-}T$_{\rm J-K}$ &  
\phantom{-}\phantom{-}${\rm BC_{V}}$ & \phantom{-}\phantom{-}${\rm M_{bol}}$ &
\phantom{-}\phantom{-}log g & & \phantom{-}T$_{\rm eff}$ &\phantom{-}\phantom{-}${\rm BC_{V}}$ &\phantom{-}\phantom{-}log g&
\phantom{-}\phantom{-}log g &\phantom{-}\phantom{-}[FeI/H] & 
\phantom{-}\phantom{-}[FeII/H] &
 \phantom{-}\phantom{-}[Fe/H] 
& ${\rm v_t }$ \\
 &  & \multicolumn{2}{c}{\hbox{2MASS}} &\multicolumn{2}{c}{\hbox{VVV}} & &  & & &  &  &    \\
 & (K) & (K) &  (K) &  (K) & (K) & & & & & (K) & & & & & & & km~s$^{-1}$  \\
\noalign{\smallskip}
\noalign{\hrule}
\noalign{\smallskip}
    2115    &  4196.6 & ---   & ---  &   ---   &    ---  & $-0.702$ &
 \phantom{-}$-0.18$ & \phantom{-}1.95 & &  4530 &\phantom{-}$-$0.457&1.99 &\phantom{-}2.00 & $-$0.98  & $-$1.02 & $-$1.00 & 1.45 \\
    2461   &  4735.2 &  --- &  --- &  4217.9 & 4506.5 & $-0.368$ &
 \phantom{-}$-0.14$ & \phantom{-}2.04 & &  4780 &\phantom{-}$-$0.323&2.04 &\phantom{-}2.05 & $-$1.13  & $-$1.09 & $-$1.11 & 1.90 \\
   2939       &  4274.4 &  4025.1 &  4423.3 &  --- &  --- & $-0.638$ &
 \phantom{-}$-0.12$ & \phantom{-}1.98 & &  4525 &\phantom{-}$-$0.440&2.01 &\phantom{-}2.00 & $-$1.07  & $-$1.07 & $-$1.07 & 1.55 \\ 
   3514     &  4496.6 &  4249.4 & 4624.0   &  4855.9  &   ---   & $-0.487$ &
 \phantom{-}\phantom{-}\phantom{-}$0.08$ & 
\phantom{-}2.09 & &  4590 &\phantom{-}$-$0.349&2.06 &\phantom{-}2.00 &
 $-$1.18  & $-$1.19 & $-$1.18 & 1.90 \\
    5037    &  4481.9 &  4254.3 &  4418.0 &  4324.0 & 4576.4  & $-0.496$
 & \phantom{-}$-0.15$ & \phantom{-}1.99 & &  4570 &\phantom{-}$-$0.429&2.05 & \phantom{-}2.15 & $-$0.98  & $-$1.03 & $-$1.0 & 1.20 \\
    5485     &  4920.4 &  4499.6 & 4197.0  &  4812.7 &  5152.4 & $-0.298$
 & \phantom{-}$0.14$ & \phantom{-}2.08 & &  4920 &\phantom{-}$-$0.282&2.09 & \phantom{-}2.07 & $-$1.18  & $-$1.18 & $-$1.18 & 1.80 \\
\noalign{\smallskip} \hline 
\end{tabular}
\end{flushleft}
\label{tabteff}
\end{table*}

\section{Photometric stellar parameters}

\subsection{Temperatures}
 The selected stars, their ID and 2MASS designations, 
 coordinates, V, I magnitudes  from Ortolani et al. (1997),
 and the 2MASS JHK$_s$ 
(Skrutskie et al. 2006)\footnote{
$\mathtt{http://ipac.caltech.edu/2mass/releases/allsky/}$}
and VVV JHK$_s$ magnitudes (Saito et al. 2012) 
are listed in Table~\ref{starmag}.
 These magnitudes and colors were used to derive initial estimates
of the effective  temperature  and  gravity, which were fine-tuned
with spectroscopic data using the  \ion{Fe}{I} \ion{and Fe}{II} lines
(see Sect. 4).

\begin{figure}
\begin{center}
\psfig{file=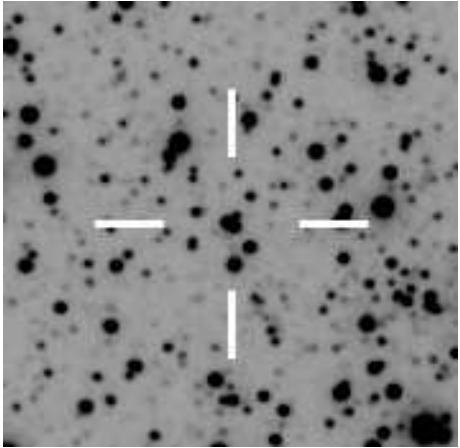,angle=0.,width=6.0 cm}
\caption{Image of star 2461 in the VVV survey, indicating
a blend of stars. Extraction of $\sim20$ arcsec.
North is $45{\circ}$ anticlockwise.
}
\end{center}
\label{image2461} 
\end{figure}

 For star 2461, the identification in the VVV
images seems to indicate a blend of at least three stars,
as can be seen in Fig. \ref{image2461}. The VVV reductions
were made in aperture photometry, which does not
yield reliable magnitudes in this case.
Only PSF photometry, which is being carried out by
the VVV group, will allow distinguishing such cases.
 For the spectroscopy, the size of 1 arcsec of the fiber does allow observing the
correct star.

The reddening value of $E(B-V)=1.19$ was derived
from $V,I$ CMDs by Ortolani et al. (1997). Barbuy et al. (1998) adopted $E(B-V)=1.21$  by using a slightly
 different R$_{\rm V}$ value.
 Barbuy et al. (2006) reported previous literature
extinction values, and based on a reanalysis of CMD data, 
adopted  $E(B-V)=1.12$. Ortolani et al. (2011)
adopted $E(V-K)=3.33$, which translates into $E(B-V)=1.21$ 
using E($V-K$)/E($B-V$)~$=2.744$ (Rieke \& Lebofsky 1985);
the use of a different reddening law does not affect
the result much.
In the present work we adopted $E(B-V)=1.12$,
following Barbuy et al. (2006).

Effective temperatures were derived from  $V-I$, $V-K$, and $J-K$ using
the color-temperature calibrations of Alonso et al.  (1999, hereafter
AAM99). These relations are similar to those by Ramirez \& Mel\'endez (2005),
as shown in their Fig. 11. The advantage of using the calibrations
of Alonso et al. is
that the bolometric corrections can be calculated, despite the disadvantage
of having to translate Cousins to Johnson I, and 2MASS to TCS JHK colors.
 To translate $V-I$ from the Cousins to the Johnson system,
 we adopted ($V-I$)$_{C}$=0.778($V-I$)$_{J}$ (Bessell 1979).
The $J,H,K_S$ 2MASS magnitudes  and colors were  translated from the 2MASS
system to  CIT (California Institute of  Technology), and from this to
TCS (Telescopio Carlos S\'anchez),  using the relations from
Carpenter (2001)  and  Alonso et  al.  (1998).  
The VVV $JHK_{s}$ colors were translated into the 2MASS $JHK_{s}$ system, using
relations reported by Soto et al. (2013).
 The derived photometric effective
temperatures are  listed in  Table~\ref{tabteff}. 

\subsection{Gravities}

The gravity values were computed using the classical formula
\[
\log g_*=4.44+4\log \frac{T_*}{T_{\odot}}+0.4(M_{\rm bol*}-M_{\rm bol{\circ}})+\log \frac{M_*}{M_{\odot}} 
\]
We adopted T$_{\odot}=5770$~K and M$_{\rm bol \odot}=4.75$ for the Sun and 
M$_{*}=0.85$~M$_{\odot}$ for the red giant branch (RGB) stars. 
For HP~1 we assumed a
 distance modulus of ($m-$M)$_{0}=14.15$ (Ortolani et al. 1997, 2011), 
and the gravities
were computed and are reported in Table~\ref{tabteff}.
Bolometric corrections were computed with formulae by 
AAM99.
 The photometric gravity values were computed assuming the bolometric
corrections from the temperature T(V-I) values (Col. 9),
and also with the spectroscopic temperature values (Col. 11).
The final spectroscopic gravities, described in the next section, 
are given in Col. 12 of Table
\ref{tabteff}, and are compatible with the photometric gravities.


\section{Spectroscopic stellar parameters}

The  equivalent  widths (EW) were measured  using  the  automatic  code
DAOSPEC, developed by Stetson \& Pancino
 (2008). We also measured EWs line by line using IRAF
for a few lines, in particular for \ion{Fe}{II} .
The EWs measured for the  \ion{Fe}{I} and \ion{Fe}{II} 
 lines are reported in Table \ref{EW}. 
We limited EWs to $20<$~EW(m{\rm \AA})~$<99$ to avoid 
on one hand,
too weak lines that are affected by the continuum level choice,
in particular when using software such as DAOSPEC, and
on the other hand, the saturated lines that are less sensitive to
abundance variations.

\begin{figure}
\centering
\psfig{file=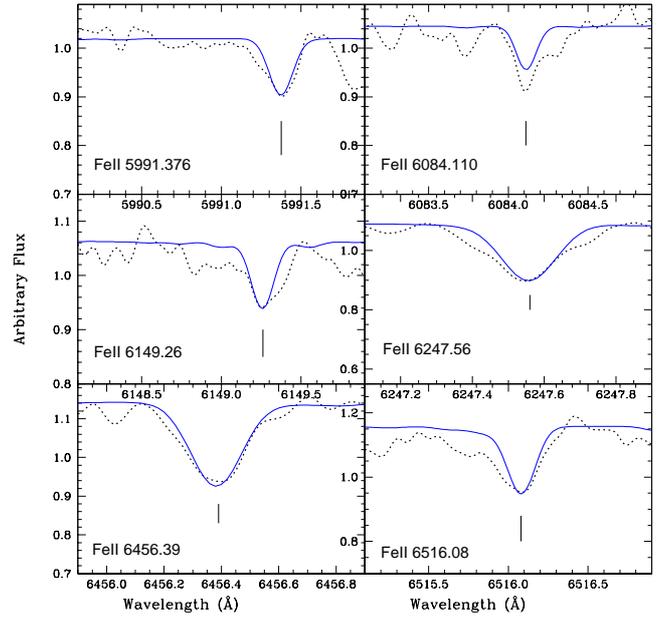,angle=0.,width=9.0 cm}
\caption{FeII lines for star 5037. Dotted lines: observed spectra.
Blue solid lines: synthetic spectra.
}
\label{fe2} 
\end{figure}

In Fig. \ref{fe2} we show the computed  \ion{Fe}{II} lines
compared with the observed spectra. Given the blends in
lines \ion{Fe}{II} 6084.11 and 6456.39 {\rm \AA}, these lines
were not used.




In the line list given in Table \ref{EW}, literature oscillator strengths 
for \ion{Fe}{I} from  
NIST\footnote{http://physics.nist.gov/PhysRefData/ASD/lines$_{-}$form.html} and
VALD3\footnote{http://vald.astro.univie.ac.at/~vald3/php/vald.php}
databases (Martin et al. 2002; Piskunov et al. 1995) are reported.
We also give the adopted values, where we have given preference to NIST over VALD.
For \ion{Fe}{II} we report the log~gf values from
Fuhr \& Wiese (2006) and Mel\'endez \& Barbuy (2009) in
Table \ref{EW}. We adopted the latter values.


Photospheric 1D models for the sample giants  were extracted from the
MARCS model atmosphere grid (Gustafsson   et al. 2008). We adopted the spherical
and mildly CN-cycled set ([C/Fe]$=-0.13$, [N/Fe]$=+0.31$).
 This choice is due to the well-known mixing that occurs along
the RGB, which transforms C into N. These models
consider [$\alpha$/Fe]=+0.20 for [Fe/H]=-0.50 and 
[$\alpha$/Fe]=+0.40 for [Fe/H]$\leq-$1.00.
The LTE  abundance  analysis and  the spectrum  synthesis calculations
were performed using the code described in
 Barbuy et al. (2003) and Coelho et al. (2005). An Fe abundance
of $\epsilon$(Fe)$=7.50$ (Grevesse \& Sauval 1998) was adopted.
Molecular lines of MgH (A$^2$$\Pi$-X$^2$$\Sigma$), 
CN  (A$^2$$\Pi$-X$^2$$\Sigma$), C$_2$ Swan (A$^3$$\Pi$-X$^3$$\Pi$), TiO
(A$^3$$\Phi$-X$^3$$\Delta$) $\gamma$   and  TiO (B$^3$$\Pi$-X$^3$$\Delta$)
$\gamma$' systems  are taken   into  account.

The stellar   parameters  were  derived    by initially  adopting  the
photometric effective  temperature  and  gravity,  and   then  by further
constraining  the  temperature by imposing  excitation equilibrium for
\ion{Fe}{I} lines.
 \ion{Fe}{I}  and  \ion{Fe}{II} lines allowed deriving gravities
by imposing ionization equilibrium. 
Microturbulence velocities v$_t$  
were  determined by canceling the trend of \ion{Fe}{I} abundance vs.  
equivalent width.

\begin{figure}
\centering
\psfig{file=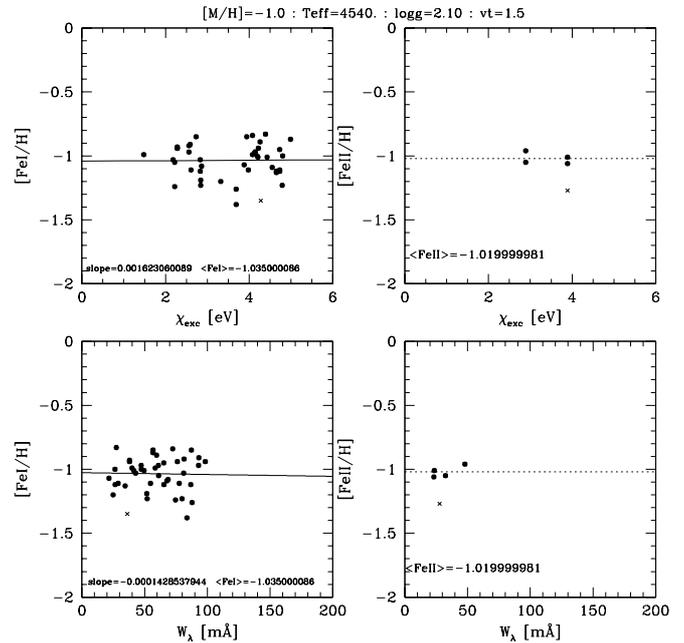,angle=0.,width=9.0 cm}
\caption{Excitation and ionization equilibria of \ion{Fe}{I} and \ion{Fe}{II} lines for star 2939.
}
\label{abon} 
\end{figure}

The final spectroscopic parameters T$_{\rm eff}$, log~g, [\ion{Fe}{I}/H],
 [\ion{Fe}{II}/H],  [Fe/H], and
 v$_t$ values  are reported in  the  last columns
  of Table~\ref{tabteff}. An example of
excitation and ionization equilibria using \ion{Fe}{I} and \ion{Fe}{II} lines
is shown in Fig. \ref{abon} for star HP1-2939.


\section{Abundance ratios}

Abundance ratios  were   obtained  by means of line-by-line  spectrum
synthesis calculations compared to the observed spectra
for the line lists given in Tables~ \ref{c2cn}, \ref{light}, 
\ref{alpha}, and \ref{heavy}.
 The fits were made by eye using a series of calculations for
different abundances and verifying the continua over 20 {\rm \AA}. We
then zoomed and fine-tuned them locally.
 The solar abundances were 
adopted from  Grevesse et al. (1998), which
are close to the latest values by
Asplund et al. (2009), Grevesse et al. (2014)
 or Lodders (2009). For oxygen we adopted
$\epsilon$(O)$=8.77$ following Allende Prieto et al. (2001) for 1D model
atmospheres, which is very close
to the value of $\epsilon$(O)$=8.76$  recommended
from 3D models by Steffen et al. (2015).

\subsection{Carbon, nitrogen, and oxygen}

\begin{table}
\begin{flushleft}
\caption{Carbon, nitrogen, and oxygen abundances derived
from C$_2$ (0,1), CN (5,1), and [OI] lines.}
\label{c2cn}      
\centering 
\small         
\begin{tabular}{l@{}r@{}r@{}r@{}r@{}r@{}r@{}rr}     
\noalign{\smallskip}
\hline\hline    
\noalign{\smallskip}
\noalign{\vskip 0.1cm} 
\hbox{line} & \phantom{-}\phantom{-}\phantom{-}\hbox{$\lambda$({\AA})} &
 \phantom{-}\phantom{-}2115& \phantom{-}\phantom{-}2641 &
\phantom{-}\phantom{-}2939 &\phantom{-}\phantom{-}3514 &
\phantom{-}\phantom{-}5037 &\phantom{-}\phantom{-}5485  &   \\
\noalign{\smallskip}
\noalign{\hrule\vskip 0.1cm}
\hbox{C$_2$(0,1)} & \phantom{-}5635.5   &\phantom{-}$\leq$+0.0   & \phantom{-}$\leq$+0.0  & \phantom{-}$\leq$+0.0 & \phantom{-}$\leq$+0.0  & \phantom{-}$\leq$+0.0 & \phantom{-}$\leq$+0.0 & \\
\hbox{CN(5,1)} & \phantom{-}6332.2     &\phantom{-}$\leq$+0.7   & \phantom{-}$\leq$+0.5  & \phantom{-}$\leq$+0.5  & \phantom{-}$\leq$+0.8  & \phantom{-}$\leq$+0.5 & \phantom{-}$\leq$+0.5 & \\
\hbox{[OI]} & \phantom{-}6300.3      &+0.4   & +0.5  & +0.5  & +0.4 & +0.35 & +0.4 & \\
\noalign{\smallskip} \hline \end{tabular}
\end{flushleft}
\end{table}

\begin{table*}
\begin{flushleft}
\caption{Abundances of light elements  Na, Mg, Al.}
\label{light}      
\centering 
\small         
\begin{tabular}{lrrrrrrrrrrrr}     
\noalign{\smallskip}
\hline\hline    
\noalign{\smallskip}
\noalign{\vskip 0.1cm} 
species & {\rm $\lambda$} ({\rm \AA}) & $\chi_{ex}$(eV) & log gf
& \hbox{2115} &\hbox{2461} &\hbox{2939} &\hbox{3514} 
&\hbox{5037} &\hbox{5485} & \hbox{HP1-2}  & \hbox{HP1-3}   \\
\noalign{\vskip 0.1cm}
\noalign{\hrule\vskip 0.1cm}
\noalign{\vskip 0.1cm}
NaI     & 5682.633 & 2.102439 & $-$0.706 &  $-$0.15 & -0.30 &$-$0.30
& +0.00 & $-$0.10 & $-$0.30 &+0.00 &$-$0.20 & \\
NaI     & 5688.194 & 2.104571 & $-$1.400 & +0.20 &-0.30 &-0.15
 &-0.20 & $-$0.10 & $-$0.30  &$-$0.10 &$-$0.20 & \\
NaI     & 5688.205 & 2.104571 & $-$0.45 & +0.20 &-0.30 &-0.15
 &-0.20 & $-$0.10 & $-$0.30 &$-$0.10 &$-$0.20 &  \\
NaI & 6154.230 & 2.102439  & $-$1.56 & 0.05 &$-$0.10 & $-$0.15
 & $-$0.15: &+0.20 & $-$0.30 &$-$0.30 &+0.05 & \\
NaI & 6160.753 & 2.104571 &  $-$1.26& +0.25 &$-$0.15 & $-$0.15
  &+0.10  & +0.30 & +0.00 &$-$0.30 &--- & \\
AlI     & 6696.185 & 4.021753 & $-$1.576 & $-$0.1 &+0.0 &+0.1 
& +0.00 &+0.20 & +0.20: &-0.25 &+0.15 &  \\
AlI     & 6696.788 & 4.021919 & $-$1.421 & --- & --- & --- 
& --- & --- & --- & --- &--- & \\
AlI     & 6698.673 & 3.142933 & $-$1.647 & $-$0.10 &+0.10 &0.00 
&+0.10 &+0.20 & +0.20 &$-$0.30 &+0.20 &  \\    
MgI & 6318.720 & 5.108171 &$-$2.10  & +0.30 &+0.30 &+0.50
 &+0.40 &  +0.40   &+0.40 &+0.15 &+0.35  & \\
MgI & 6319.242 & 5.108171 &$-$2.36 & +0.30 &+0.40 &+0.30 
&+0.30 & +0.20   &+0.40  &--- &+0.30 & \\
\noalign{\smallskip} \hline \end{tabular}
\end{flushleft}
\end{table*}

\makeatletter
\renewcommand{\table}[1][]{\@float{table}[!htp]}
\makeatother 

\begin{figure}
\centering
\caption{[Na/Fe] vs. [O/Fe] for the sample stars compared
with stars of NGC 6121. Symbols: 
present work: green filled squares: the 6 HP~1 stars with [Fe/H]~$=-1.0\pm0.05$;
 open green squares: the 2 HP~1 stars with [Fe/H]=-1.18,
compared with blue filled triangles: stars of NGC 6121.  }
\label{na-o} 
\psfig{file=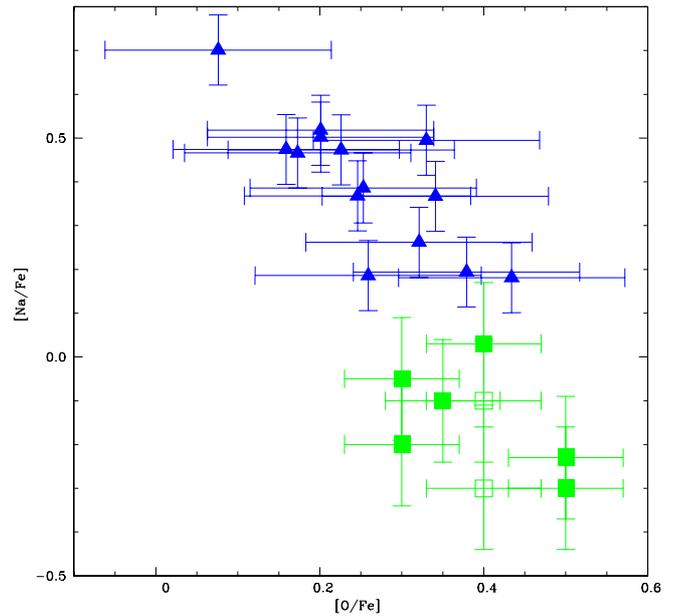,angle=0.,width=9.0cm}
\end{figure}

The carbon abundances were estimated from the C$_{2}$(0,1) bandhead
at 5635.3 {\rm \AA}.  The list of laboratory C$_2$ lines (Phillips \&
Davis 1968) was reported in Barbuy et al. (2014). Because
the feature is weak in these metal-poor stars, we
adopted [C/Fe]$=0.0$ and just checked that this was a suitable
upper limit. 
The nitrogen abundances were measured using the CN (5,1) 6332.18 {\rm \AA}
 of the CN A$^{2}\Pi$-X$^{2}\Sigma$ red system. The CN feature
is also weak, but can be used to estimate a reliable upper limit for the
N abundance.

The forbidden oxygen [OI] 6300.311 {\rm \AA}
line was used to derive the oxygen abundances. 
A check for telluric lines was carried out
by overplotting the spectrum of a fast-rotation B star,
and we verified that none of the stars were affected by them.
 The resulting C, N, and O abundances are given in Table
\ref{c2cn}, showing the essentially adopted  carbon abundance
of [C/Fe]$\sim0.0$, the enhanced nitrogen
abundances as expected in giants, and enhanced oxygen
abundances of $+0.35<$~[O/Fe]~$<+0.5$, 
typical of enrichment by SNII.

\subsection{Odd-Z elements Na, Al, and alpha-elements}

In Table \ref{light} and \ref{alpha} we report the line-by-line
abundances of the odd-Z elements Na, Al, and
the alpha-elements Mg, Si, Ca, and Ti abundances.
We have inspected the abundance results as a function
of effective temperature and microturbulence velocity
and found no trend for any of the elements.
In addition to the six sample stars, 
we also rederived the abundances
of Na, Al and Mg for the two stars analyzed in 
Barbuy et al. (2006), as given in Table 6.



\begin{table*}
\caption{Abundance ratios [X/Fe] of alpha-elements O, Mg, Si, Ca, Ti,
 and atomic parameters.
}
\label{alpha}
\begin{flushleft}
\small
\begin{tabular}{llllrrrrrrrrrrrrrrrrr}
\noalign{\smallskip}
\hline\hline    
\noalign{\smallskip}
\noalign{\vskip 0.1cm} 
Species & \hbox{\rm $\lambda$ ({\rm \AA})} & \hbox{\rm $\chi_{ex}$ (eV)} 
&\hbox{log~gf} & {2115}  &   {2461} & {2939} & {3514} &  {5037} & {5485} &  \\
\noalign{\smallskip}
\hline\hline    
\noalign{\smallskip}
\noalign{\vskip 0.1cm} 
SiI & 5665.555 & 4.920417 &$-$2.04 & +0.1 & +0.2 &+0.2
 & +0.3 &0.0 & +0.2 &  \\ 
SiI & 5666.690 & 5.616073 & $-$1.74 &  --- & --- & ---
 &+0.3 & +0.1 & +0.2 & \\
SiI & 5690.425 & 4.929980 & $-$1.87 &+0.3 & +0.2 &+0.3
 & +0.3 & +0.2 & +0.0 &  \\
SiI & 5948.545 & 5.082689 & $-$1.30 &+0.4 & +0.4 & +0.5 
& +0.2 & +0.3 & +0.2 & \\
SiI & 6142.494 & 5.619572   & $-$1.50 & +0.3 & +0.5 &+0.25
 & --- & +0.2 & +0.1 & \\
SiI & 6145.020 & 5.616073  & $-$1.45 & +0.3 &+0.3 &+0.2
   & ---  &+0.1 & +0.3 & \\
SiI & 6155.142 & 5.619572  & $-$0.85 & +0.2 &+0.2 & +0.4   
  & +0.4 & +0.2 & --- & \\
SiI & 6237.328 & 5.613910   &  $-$1.01 & +0.2 & +0.2 & +0.3
   & +0.2 & +0.2   &+0.1 & \\
SiI & 6243.823 & 5.616073   & $-$1.30 & +0.2 & --- & +0.3   &
 +0.4 &+0.1  &  --- \\
SiI & 6414.987 & 5.871240    & $-$1.13 &+0.3& +0.4 & +0.4 
 & +0.3  &+0.4 & +0.2 &  \\
SiI & 6721.844 & 5.862872  & $-$1.17 &+0.2 & +0.20 & +0.4   
&+0.4 & +0.2  &+0.0 \\
CaI     & 5601.277 & 2.525852 & $-$0.52 &  +0.0 & +0.0 &+0.0
 &$-$0.1 &+0.0 &$-$0.3 \\
CaI     & 5867.562 & 2.932710 & $-$1.55 &  +0.1 & +0.1 &+0.0
 & +0.3 &+0.0 &  --- \\
CaI     & 6102.723 & 1.879467 & $-$0.79 &  +0.3 & +0.1 &+0.1
 & $-$0.2 & +0.2 & +0.0 & \\
CaI     & 6122.217 & 1.885935 & $-$0.20 &  +0.0 & +0.1 &+0.0
 & +0.0 & +0.2 &+0.1  \\
CaI & 6161.295 & 2.523157   & $-$1.02& +0.3 & +0.1 &+0.4   
 & +0.2   & +0.2 & +0.0     \\
CaI & 6162.167 & 1.899063   & $-$0.09& +0.3 & +0.1 &+0.1
 & +0.1 & +0.2 & +0.0 & \\
CaI & 6166.440 & 2.521433  & $-$0.90 &  +0.3 &+0.0 &+0.4  
 & +0.1 & +0.3   & +0.0    & \\
CaI & 6169.060 & 2.523157  & $-$0.54&   +0.3   & +0.1 & +0.1  
 & +0.2 & +0.4   &+0.1&\\
CaI & 6169.564 & 2.525852   & $-$0.27 &  +0.3   &+0.1 &+0.2  
 & +0.0 & +0.2  &+0.0 & \\
CaI & 6439.080 & 2.525852   & +0.30  & +0.3   & +0.0 &+0.1
   & +0.0 & +0.4   &+0.1 \\
CaI & 6455.605 & 2.523157  & $-$1.35     &  +0.3   & +0.1 &+0.4
   & +0.2 & +0.4   &+0.0    \\
CaI & 6464.679 & 2.525852   &$-$2.10   & +0.4&+0.4 &\phantom{-}$>$+0.5
   & --- & +0.2  & --- \\
CaI & 6471.668 & 2.525852  & $-$0.59&   +0.3 &+0.2 &+0.3
 & +0.2 & +0.4 &+0.3 &    \\
CaI & 6493.788 & 2.521433   & \phantom{-}0.00  &  +0.3 & $-$0.1 &+0.1
 &  +0.0 & +0.3  &  +0.0 & \\
CaI & 6499.654 & 2.523157     & $-$0.85  & +0.3 & +0.0 & +0.3
   & +0.1 & +0.2  & +0.0 & \\
CaI & 6572.779 & 0.00     & $-$4.32& +0.5 & 0.0 &+0.3
   & +0.2 & +0.3 & +0.0 &\\
CaI & 6717.687 & 2.709192      & $-$0.61  & +0.4 &+0.2 &+0.5
   & +0.3 &+0.5 & +0.1 \\
TiI     & 5689.459 & 2.296971 & $-$0.47  &+0.2 & --- & ---
      & +0.1 & ---  & --- &  \\
TiI     & 5866.449 & 1.066626 & $-$0.84  &+0.1 & +0.0 & +0.2
& +0.0 & +0.3  &  ---   & \\ 
TiI     & 5922.108 & 1.046078 & $-$1.46  &+0.3 &+0.2 &+0.4
  &  +0.1 & +0.0 & ---- &  \\
TiI     & 5941.750 & 1.052997 & $-$1.53  &+0.2 &+0.2 &+0.3
  &+0.2 & +0.2 &  +0.2 \\
TiI     & 5965.825 & 1.879329 & $-$0.42  &+0.3 &+0.2 &+0.3 
& +0.1 & +0.1 & +0.0 &  \\
TiI     & 5978.539 & 1.873295 & $-$0.53  &+0.3 & +0.1 &+0.2
   &  +0.2  & +0.2 &+0.0 &  \\
TiI     & 6064.623 & 1.046078 & $-$1.94 &+0.3 & +0.2 &+0.4
  & +0.2  & +0.3 & --- &  \\
TiI     & 6091.169 & 2.267521 & $-$0.42 &+0.3 &+0.2 &+0.3 
& +0.3 & +0.0 &+0.3 &  \\
TiI     & 6126.214 & 1.066626 & $-$1.43 &+0.3 & +0.2 &+0.3
  &+0.2  & +0.2 &+0.0 &  \\
TiI & 6258.110 & 1.443249& $-$0.36  &+0.1  & +0.0 &+0.3
& +0.0  &+0.1 & $-$0.1 & \\
TiI & 6261.106 & 1.429852 & $-$0.48  &+0.3 &+0.1 &+0.2
   &+0.0 & +0.0 &$-$0.1 &  \\
TiI & 6303.767 & 1.443249& $-$1.57  &+0.3 & +0.15 &+0.35 &
 +0.3 &  +0.2 & +0.2 & \\
TiI & 6312.240 & 1.460236 & $-$1.60  &+0.3 &+0.3 &+0.4
 & +0.2 & +0.1 & --- &  \\
TiI & 6336.113 & 1.443249 & $-$1.74&+0.1 & --- & +0.2 &
+0.2 & +0.0 & --- &  \\
TiI & 6508.150 & 1.429852     & $-$2.05  &+0.2 &+0.20 &+0.2
 &+0.3 &+0.2 & --- &  \\
TiI & 6554.238 & 1.443249& $-$1.22  &+0.1 &+0.0 &+0.2
 & +0.2 &+0.1 & $-$0.15   \\
TiI & 6556.077 & 1.460236     & $-$1.07  &+0.2 &+0.2 &+0.3
 & +0.1   &+0.1 & +0.0  \\
TiI & 6599.113 & 0.899612     &$-$2.09 &+0.3 &+0.2 &+0.4
   & +0.2 & +0.5 & +0.3 \\
TiI & 6743.127 & 0.899612     & $-$1.73  & +0.3 &+0.1 &+0.3
 &+0.3 & +0.1 & +0.2  \\
TiII & 5418.751 & 1.581911 &$-$2.13 &+0.4 & +0.0 &+0.1
 &+0.0 & +0.1 &+0.0 &  \\ 
TiII& 6491.580 & 2.061390     &$-$2.10   &+0.4 & +0.2 &+0.3
   &+0.3 &  +0.5 &+0.1 \\
TiII& 6559.576 & 2.047844 &$-$2.35   &+0.4 &+0.3 &+0.3
   & +0.2 & +0.5   & +0.3 \\
TiII& 6606.970 & 2.061390     &$-$2.85   & +0.4 &+0.2 &+0.2 
  & +0.25 &+0.1 &+0.1 & \\
\noalign{\smallskip} 
\hline 
\end{tabular}
\end{flushleft}
\end{table*}


\subsection{Heavy elements}
\label{Sect:heavy}

\begin{figure}
\centering
\psfig{file=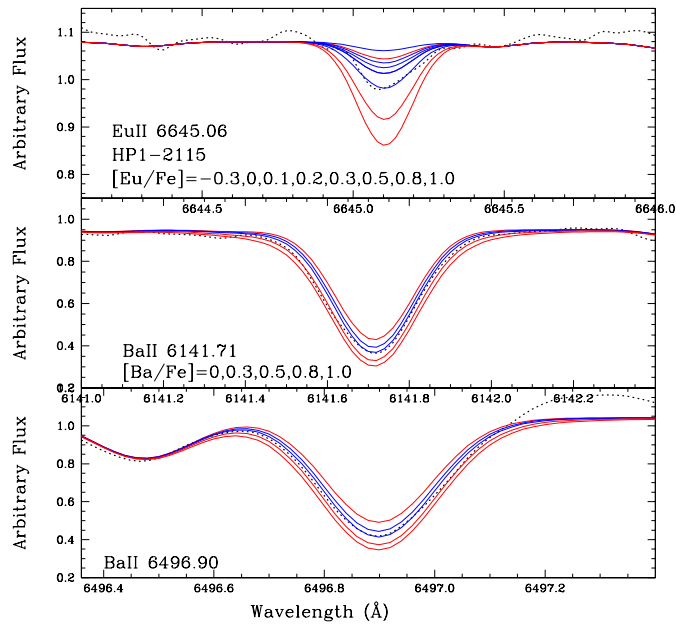,angle=0.,width=9.0 cm}
\caption{Fits to the \ion{Eu}{II} 6645,
\ion{Ba}{II} 6141 and \ion{Ba}{II} 6496 {\rm \AA} lines
 in star 2115. The [Eu/Fe] and [Ba/Fe] values adopted for the
different calculations are indicated in the panels.
}
\label{eu} 
\end{figure}

In Table \ref{heavy} we report the line-by-line derivation
of abundances for lines  of the neutron-capture dominant s-elements 
Sr, Y, Zr, La, Ba, and the r-element Eu.
As in Table \ref{light}, we also rederived the abundances
of heavy elements for the two stars analyzed in Barbuy et al. (2006).
The hyperfine structure (HFS) for the studied lines of 
\ion{La}{II}, \ion{Ba}{II} and \ion{Eu}{II} were taken into account,
as described in Barbuy et al. (2014).
The fits to the lines of \ion{Eu}{II} 6645,
\ion{Ba}{II} 6141 and \ion{Ba}{II} 6496 {\rm \AA} lines
 in star 2115 are shown in Fig. \ref{eu}.

\begin{table*}
\begin{flushleft}
\caption{Central wavelengths, log gf and abundances of heavy elements.}
\label{heavy}      
\centering 
\small         
\begin{tabular}{lrrrrrrrrrrrrrrrrr}     
\noalign{\smallskip}
\hline\hline    
\noalign{\smallskip}
\noalign{\vskip 0.1cm} 
species & {\rm $\lambda$} ({\rm \AA}) & \phantom{-}{\rm $\chi_{ex}$ (eV)} & 
\hbox{gf$_{adopted}$} & {2115} & {2461}
& \hbox{2939} &\hbox{3514} &\hbox{5037} &\hbox{5485} &\hbox{HP1-2} &\hbox{HP1-3}   \\
\noalign{\vskip 0.1cm}
\noalign{\hrule\vskip 0.1cm}
\noalign{\vskip 0.1cm}
EuII   & 6645.064 & 1.379816 & +0.12 & +0.50 & +0.50 & +0.70
 & +0.60 & +0.60 & +0.55 &+0.40  &+0.55 & \\
BaII & 6141.713 & 0.703636 & $-$0.076 & +0.50 & +0.30 & +0.50
& $-$0.10 & +0.70 & +0.50 &+0.50 &+0.50 & \\ 
BaII & 6496.897 & 0.604321 & $-$0.32 & +0.50 & +0.30 & +0.50
 &+0.10 & +0.80 & +0.40 &+0.80  &+0.45 & \\
LaII & 6262.287  & 0.403019 & $-$1.60 & +0.40 & +0.30  &+0.50
 & +0.30 &+0.10 &   --- & --- &--- & \\
LaII & 6320.376  & 0.172903 & $-$1.56  & +0.30  & +0.00  &+0.30
 &+0.25 & +0.00 & --- &$-$0.30 &$-$0.15 & \\
LaII & 6390.477  & 0.321339 & $-$1.41  & +0.30  & +0.40 &+0.50
 &+0.40 & +0.10 &  +0.10 &0.00 &$-$0.20 & \\
YI & 6435.004     & 0.065760  & $-$0.82 &+0.20 & +0.30 &+0.20
 & +0.50 & +0.20: & +0.50 &$-$0.15 & --- & \\
YII & 6795.414    & 1.738160 & $-$1.19 & +0.1 & +0.30 &+0.30
 & +0.20 & +0.00 & ---  &$-$0.15 &$-$0.10 & & \\
ZrI  & 6127.475   & 0.153855 & $-$1.18 & +0.10 & +0.00 &+0.40 
& +0.40 & +0.20 & --- &--- &+0.10 & \\
ZrI  & 6134.585   & \phantom{+}0.00 & $-$1.43 &+0.10 & +0.50 &+0.30
 & --- & +0.20  &--- &+0.20: & 0.00 & \\
ZrI  & 6140.535   & \phantom{+}0.00 & +0.10 & --- & +0.50 &
--- & --- & ---  &(+0.70) &--- &  \\
ZrI  & 6143.252   & \phantom{+}0.070727 & $-$1.50 & +0.10 & +0.10 &+0.40
 & +0.00 & +0.00 & +0.30: &--- &$-$0.15 & \\
SrI & 6503.989 & 2.258995   & +0.26   &+0.30 &--- & +0.30: 
& --- & +0.30:: & --- &+0.50 &+0.30: & \\ 
SrI & 6791.016 & 1.775266   & $-$0.73 &+0.80: & --- & --- 
& --- & +0.50:: & --- &--- &--- & \\
\noalign{\vskip 0.1cm}
\noalign{\hrule\vskip 0.1cm}
\noalign{\vskip 0.1cm}  
\hline                  
\end{tabular}
\end{flushleft}
\end{table*}

\subsection{Errors}

 The errors due to uncertainties in spectroscopic parameters are given 
in Table \ref{errors}, applied to the sample star HP~1: 2115.
The error  on the slope in the  FeI $vs.$ excitation potential implies
an error in  the temperature of $\pm100$~K for the sample  
stars. An uncertainty of the order of $0.2$~km~s$^{-1}$ on the microturbulence
velocity is estimated from the imposition of a constant value of [Fe/H] as
a function of EWs. Errors based on EWs are given on FeI and FeII abundances.
 
The errors on the abundance ratios [X/Fe] were computed by fitting the lines
with the modified atmospheric model. 
The error reported corresponds to the new value obtained by using
the modified model atmosphere. 
The element abundance ratios, induced  by   a change of  $\Delta$T${\rm
eff}=+100$~K, $\Delta$log~g$=+0.2$, $\Delta$v$_{\rm  t}=0.2$~km~s$^{-1}$, 
and a total error estimate, are given in Table \ref{errors}.
 These errors are overestimated because the stellar 
parameters are covariant. The correlation matrix is
difficult to estimate, however, and would add other error sources, so that
we preferred the quadratic sum of the diagonal terms as reliable.

 Additionally, an uncertainty of about 0.8 m{\rm \AA}
in the EWs of the Fe lines was estimated with the formula of 
(Cayrel 1988; Cayrel 2004).  
With a mean FWHM = 12.5 pixels, or 0.184 {\rm \AA},
a CCD pixel size of 15 $\mu$m, or $\delta{x}$ = 0.0147 {\rm \AA}
in the spectra, and assuming a mean
S/N=100, we obtain an error $\Delta$EW $\sim$ 0.8 m{\rm \AA}.

We derived abundances uniquely from fitting synthetic spectra,
such that we need to take the 
uncertainty in the continuum placement into account. We estimate an error of
0.1dex for the weak and strong lines and 0.05 dex for 
medium lines. This is included in Table \ref{errors}.

\begin{table}
\caption{Abundance uncertainties for star 2115
 for uncertainties of $\Delta$T$_{\rm eff}=+100$~K,
$\Delta$log~g$=+0.2$, $\Delta$v$_{\rm t}=0.2$~km s$^{-1}$,
an assumed error in EWs or continuum placement,  and
the corresponding total error. The errors correspond to the
difference in abundance obtained with the modified parameters.} 
\label{errors}
\begin{flushleft}
\small
\tabcolsep 0.15cm
\begin{tabular}{lccccc}
\noalign{\smallskip}
\hline
\noalign{\smallskip}
\hline
\noalign{\smallskip}
\hbox{Abundance} & \hbox{$\Delta$T} & \hbox{$\Delta$log $g$} & 
\phantom{-}\hbox{$\Delta$v$_{t}$} &
\phantom{-}\hbox{EW} 
& \phantom{-}\hbox{($\sum$x$^{2}$)$^{1/2}$} \\
\hbox{} & \hbox{100 K} & \hbox{0.2 dex} & \hbox{0.2 kms$^{-1}$} & \\
\hbox{(1)} & \hbox{(2)} & \hbox{(3)} & \hbox{(4)} & \hbox{(5)} \\
\noalign{\smallskip}
\hline
\noalign{\smallskip}
\noalign{\hrule\vskip 0.1cm}
\hbox{[FeI/H]}       &  +0.06     &  +0.03   & $-$0.07  &0.02  &\phantom{+}0.10 \\
\hbox{[FeII/H]}      & $-$0.12    &  +0.11   & $-$0.03  &0.02  &\phantom{+}0.17  \\
\hbox{[N/Fe]}        &  $-$0.05   & +0.00    &  +0.00   &0.10  &\phantom{+}0.11  \\
\hbox{[O/Fe]}        &  +0.00     & $-$0.05  & +0.00    &0.05  &\phantom{+}0.07  \\
\hbox{[NaI/Fe]}      & +0.10      & +0.00    & $-$0.01  &0.10  &\phantom{+}0.14  \\
\hbox{[Al/Fe]}       & +0.05      &+0.00     & +0.00    &0.10  &\phantom{+}0.11  \\
\hbox{[MgI/Fe]}      & +0.03      &  +0.00   & +0.00    &0.10  &\phantom{+}0.10 \\
\hbox{[SiI/Fe] }     & +0.00      &  +0.02   & $-$0.02  &0.05  &\phantom{+}0.06  \\
\hbox{[CaI/Fe]}      &  +0.10     & +0.00    & $-$0.05  &0.05  &\phantom{+}0.12  \\
\hbox{[TiI/Fe]}      &  +0.10     &  +0.00   & $-$0.03  &0.05  &\phantom{+}0.12  \\
\hbox{[TiII/Fe]}     & $-$0.10    &  +0.08   & $-$0.01  &0.05  &\phantom{+}0.14  \\
\hbox{[SrI/Fe]}      & $-$0.15    & +0.00    & +0.00    &0.20  &\phantom{+}0.25  \\
\hbox{[YI/Fe]}       & $-$0.15    &  +0.00   &  +0.00   &0.10  &\phantom{+}0.18  \\
\hbox{[YII/Fe]}      &  +0.02     &  $-$0.15 &  +0.00   &0.10  &\phantom{+}0.18 \\
\hbox{[ZrI/Fe]}      &  $-$0.20   & +0.00    &  +0.00   &0.10  &\phantom{+}0.22  \\
\hbox{[BaII/Fe]}     &  +0.02     &  +0.02   &$-$0.20   &0.10  &\phantom{+}0.23  \\
\hbox{[LaII/Fe]}     &  +0.00     & +0.03    & +0.00    &0.10  &\phantom{+}0.10  \\
\hbox{[EuII/Fe]}     & +0.00      & +0.03    & +0.00    &0.10   &\phantom{+}0.10  \\
\noalign{\smallskip} 
\hline 
\end{tabular}
\end{flushleft}
\end{table}


\section{Discussion}
In Table \ref{meanabundances} we report the mean abundances for
each star and for each element, and in the last line we report
the radial velocities.

\subsection{Two stellar populations?}

We derived a mean metallicity of  [Fe/H]$=-1.06\pm0.15$ by adding
the two stars previously analyzed in HP~1 by Barbuy et al. (2006).

Two stars, 3514 and 5485, show a slightly lower metallicity of
[Fe/H]$=-1.18$ and, possibly not by coincidence, they
also show radial velocities of 
 v$^{\rm hel}_{\rm r}=+35.6$ and $+34.7$~km~s$^{\rm {-1}}$,
 respectively,
which is lower than the radial velocities of the other six stars, whose
values lie in the range  $-41.0<$~v$^{\rm hel}_{\rm r}<-47.0$~km~s$^{\rm {-1}}$. 
Their radial velocities are compatible with
being members within the uncertainties.
If these two stars correspond to an earlier stellar generation
in the cluster, then
we might have a first stellar
generation with [Fe/H]=$-$1.18
and a second generation with a mean metallicity of [Fe/H]$=-1.02\pm0.05,$
as found for the other six stars.

The membership of these two lower velocity stars is another question.
Recent work by Bellini et al. (2015) for NGC 2808 has shown differences
in radial velocities between
two stellar populations in a cluster. Malavolta et al. (2015)
studied M4, showing that it has a dispersion of 4 km~s$^{\rm {-1}}$. HP~1 is probably subject
to tidal effects that lead to possible perturbations
or even disruption. The two low-velocity stars might become unbound from the cluster,
given their difference in velocites of 
$\Delta$v$^{\rm hel}_{\rm r}$=$-$10 km~s$^{\rm {-1}}$ .


\begin{table*}
\caption[1]{Mean abundances of C, N, odd-Z elements Na, Al,
and $\alpha$-elements O, Mg, Si, Ca, Ti, and heavy elements
Y, Sr, Zr, Ba, La, and Eu. Mean values for NGC 6522 are given in the
last column. }
\begin{flushleft}
\tabcolsep 0.15cm
\begin{tabular}{ccccccccccccccc}
\noalign{\smallskip}
\hline
\noalign{\smallskip}
\hline
\noalign{\smallskip}
{\rm [X/Fe]} & & 2115 & 2461 & 2939 & 3514 & 5037 & 5485 & HP1-2 & HP1-3 &
& \hbox{Mean} & NGC 6522 \cr
\noalign{\vskip 0.2cm}
\noalign{\hrule\vskip 0.2cm}
\noalign{\vskip 0.2cm}
C & & $\leq$+0.00   &$\leq$+0.00   &$\leq$+0.00   &$\leq$+0.00   &$\leq$+0.00 &$\leq$+0.00   &+0.00   &+0.00 &&+0.00&$-$0.03&  \cr
N & & $\leq$+0.70   &+$\leq$0.50   &$\leq$+0.50   &$\leq$+0.80   &$\leq$+0.50 &$\leq$+0.50   &+0.50   &+0.20 &&+0.53&+0.67&  \cr
O & & +0.40   &+0.50   &+0.50   &+0.40   &+0.35 &+0.40   &+0.30   &+0.30 &&+0.40&+0.36& \cr
Na & &+0.03   &$-$0.30 &$-$0.23 &$-$0.10 &$-$0.10 &$-$0.30 &$-$0.05  &$-$0.20&&$-$0.16&+0.05&  \cr
Al & &$-$0.15 &+0.05 &+0.05 &+0.08   &+0.20 &+0.20   &$-$0.28  &+0.18 &&+0.04&+0.20&  \cr
Mg & &+0.30   &+0.35   &+0.65   &+0.35   &+0.33 &+0.40   &+0.15  &+0.33&&+0.36&+0.23&  \cr
Si & &+0.25   &+0.31   &+0.33   &+0.31   &+0.20 &+0.15   &+0.30  &+0.30&&+0.27&+0.13&  \cr
Ca & &+0.28   &+0.11   &+0.21   &+0.11   &+0.26 &+0.04   &$-$0.04&+0.10&&+013&+0.13&  \cr
TiI & &+0.26   &+0.16   &+0.28   &+0.17   &+0.16 &+0.06   &+0.07  &+0.08&&+0.16&+0.04&  \cr
TiII& &+0.40    & +0.18   &+0.23  &+0.19   &+0.31  &+0.13 &+0.10 &+0.15 &&+0.21&+0.17& \cr
Y  & &+0.15   &+0.30   &+0.25   &+0.35   &+0.10 &+0.50   &-0.15  &$-$0.1 &&+0.18&+0.31&  \cr
Zr & &+0.15   &$-$0.10 &+0.28   &+0.37   &+0.20 &+0.15   & ---   &$-$0.02 &&+0.15&+0.18&  \cr
Sr & &+0.55:  &---     &+0.30   &---     &+0.40 &---     &+0.50: &+0.30:&&+0.41:&+0.23&  \cr
Ba & &+0.50   &+0.30   &+0.50   &+0.00   &+0.75 &+0.45   &+0.65 &+0.47 &&+0.46&+0.32&  \cr
La & &+0.33   &+0.23   &+0.43   &+0.32   &+0.08 &+0.10   &$-$0.15&$-$0.18 &&+0.23& ---  \cr
Eu & &+0.50   &+0.50   &+0.70   &+0.60   &+0.60 &+0.55   &+0.40  &+0.55 &&+0.15&+0.30&  \cr
Ba/Eu& & 0.0  &$-$0.20   &$-$0.20   &$-$0.60   &+0.15 &$-$0.10   &+0.25  &$-$0.08 &&+0.31& --&\cr
\noalign{\vskip 0.2cm}
\noalign{\hrule\vskip 0.2cm}
\noalign{\vskip 0.2cm}
v$_{\rm r}$~(km~s$^{\rm {-1}}$) &  &+41.0 &+41.0 &+46.0 &+35.0 &+41.0 &+34.0 &+44.6 &+44.0 && & & \cr
\noalign{\smallskip} \hline \end{tabular}
\label{meanabundances}
\end{flushleft}
\end{table*}

\subsection{Odd-Z elements and Na-O anticorrelation}

 Na and Al tend to be underabundant in HP~1. This may
be due to internal nucleosynthesis (Gratton et al. 2012).
In particular, 
 a Na-O anticorrelation due to the Ne-Na cycle of 
proton capture reactions was shown by Carretta et al.
(2009) to be confirmed for several globular clusters. 
In Fig. \ref{na-o} we show the [Na/Fe] vs. [O/Fe]
 for 14 stars of NGC 6121 by Carretta et al. (2009),
compared to the present results for HP~1.
This figure points to  evidence for a Na-O anticorrelation
in HP~1, but with the difference with
respect to NGC 6121 and  most other
clusters studied by Carretta et al. (2009), 
of Na being lower than in the other clusters.
Na is originally probably underabundant in HP~1, which reinforces
its peculiar pattern.
 
\subsection{Alpha-elements}

The $\alpha$-element enhancements in  O, Mg, and Si together with the enhancement of 
the r-process element Eu are indicative of a fast early 
enrichment by SNII.
Ca and Ti are only slightly enhanced with [Ca/Fe]$=+0.13$ 
and [Ti/Fe]$=+0.18$ (a mean of \ion{Ti}{I} and \ion{Ti}{II}
abundances).
The difference between O, Mg, and Si on one hand and Ca and Ti on the
other was also detected in other bulge samples, such as NGC 6522 (Barbuy et al. 2014). The mean values for NGC 6522 are reported for comparison purposes in Table 10.

The fact that [O,Mg,Si/Fe] are higher than [Ca,Ti/Fe]
does not match the results for field stars by Gonz\'alez
et al. (2011). This might be explained by the mass of 
supernovae type II that enriched the cluster, or else
by the absence of contribution from supernovae type I.

\subsection{Heavy elements}

\begin{figure*}
\centering
\caption{[Y,Zr,Ba,La,Eu/Fe] vs. [Fe/H] and [Y/Ba] vs. [Fe/H] for the sample stars compared
with literature values. Symbols:
 blue filled squares: NGC 6522; magenta pentagons: M62;
red filled triangles: bulge field dwarfs by Bensby et al.
(2013);
blue triangles: bulge field red giants from Rich et al. (2012);
cyan triangles: Van der Swaelmen et al. (2016);
yellow triangles: Siqueira-Mello et al. (2016);
green filled pentagons: HP~1 from the present work.
 }
\label{plotheavy} 
\psfig{file=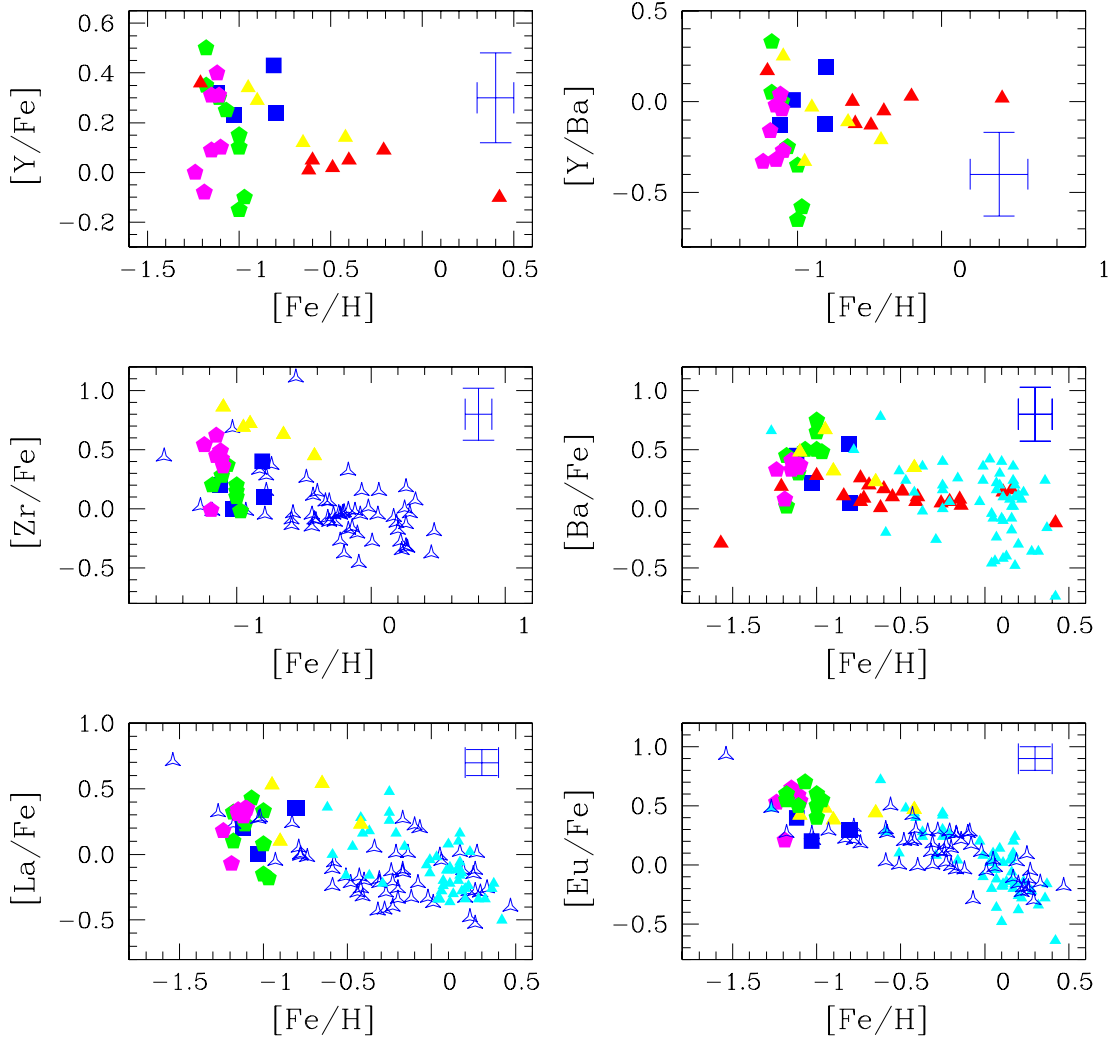,angle=0.,width=15.0cm}
\end{figure*}

Figure \ref{plotheavy} shows the results for
Y, Zr, Ba, La, and Eu compared to other available
heavy element abundance determinations in bulge stars.
The Sr abundances are not very reliable because the
lines used are extremely faint. We therefore did not plot this element.
Literature data include
a) field red giants in Plaut's field analyzed by Johnson et al. (2012),
b) seven red giants in the globular cluster M62 (NGC 6266)
by Yong et al. (2014),
c) 
microlensed bulge dwarf stars analyzed by Bensby et al. (2013),
d) four red giants in NGC 6522 analyzed by Barbuy et al. (2014),
e) 56 bulge field red giants for which Van der Swaelmen et al. (2016)
derived heavy element abundances, and
f) five bulge field red giants with [Fe/H]$\approx-1.0$ from the
sample by Ness et al. (2013a) that were analyzed by Siqueira-Mello
et al. (2016).


This figure shows that 
 [Zr/Fe], [Ba/Fe], [La/Fe], and [Eu/Fe] increase steadily 
with decreasing metallicity.
 At a metallicity of around [Fe/H]$\approx-1.0$, the dominantly s-elements Y, Zr, Ba, and La show an abundance spread.
 This behavior is compatible with expectations from massive spinstars:
a spread like this is predicted from the models by 
Frischknecht et al. (2016, and references therein) 
and Meynet et al. (2015, and references therein), as discussed and shown
in Chiappini et al. (2011), Chiappini (2013), and Barbuy et al. (2014).

On the other hand, the enhancements of O, Mg, and Eu, and to a lesser
extent of Si, Ca, and Ti, indicate an early enrichment by supernovae
type II. The small fraction of r-process production of the dominantly 
s-elements in the solar neighborhood might be responsible for 
their production at early times, as first suggested by Truran (1981).
To verify the r- or s-nature of the
heavy element abundances, we report the [Ba/Eu] ratios in penultimate 
line of Table \ref{meanabundances}.
 
The ratio [Ba/Eu] is indicative of the relative contribution
of the s- and
 r-process.
We find $-0.60<$~[Ba/Eu]~$<+0.25$, and [Ba/Eu]$_{\rm r}\sim-0.8$ is
given as typical of a pure r-process by  Bisterzo et al. (2014).
This means that the barium-to-europium ratio is above the line for a pure
r-process and might be interpreted as due to an s-process.
The s-process can be due to transfer of matter from an AGB companion
 or  to a general AGB pollution, as recently made
likely to occur in multi-population clusters
(e.g. Renzini et al. 2015), or to spinstars. 
The spread in abundances of Y, Sr, Zr, Ba, and La
at around [Fe/H]$\sim-1.0$ is the main indicator
of the contribution by spinstars (Chiappini et al. 2011). 
Finally, both r- and s- processes may take place, that is, massive spinstars producing s-elements may not exclude a later explosion of the supernova and the subsequent r-process. 
Another possible explanation would be an additional process that enhanced 
the lightest heavy elements at early times. Several models are 
proposed in the literature, such as the lighter element primary process LEPP.
 Travaglio et al. (2004) discussed this process for the Sun. A different LEPP 
mechanism was discussed by Montes et al. (2007) for metal-poor stars.
A possible explanation for the LEPP is the weak r-process 
(Wanajo \& Ishimaru 2006) or supernovae neutrino-driving winds
(Arcones \& Thielemann 2013). 
 We note that different processes
might be enriching in Y, Sr, and Zr at different metallicities and environments.
 Niu et al. (2015) proposed a unified solution.
Finally, it is important to mention that, as pointed out by
Roederer et al. (2010) (see his Fig. 11), there are varying degrees
of enrichment of first- and second-peak elements
 in stars enriched by neutron capture. 

\section{Conclusions}

We carried out a detailed analysis of six red giants of the
bulge moderately metal-poor globular cluster  HP~1 and added
two other previously analyzed stars.
A metallicity of [Fe/H]$=-1.06\pm0.15$ was derived from the eight
stars, and since HP~1 has a blue  horizontal branch, this 
combination of characteristics is an indication of a very old age.

We found overabundances of the 
$\alpha$-elements [Mg/Fe]$\approx$[O/Fe]$\approx$+0.4 and 
[Si/Fe]$\approx$+0.3 and lower values of
[Ca/Fe]$\approx$[Ti/Fe]$\approx$+0.10.
The light odd-Z elements Na and Al are low with
[Na/Fe]=$-$0.20 and [Al/Fe]=+0.18.
 In particular, because of the low Na abundance, 
the Na-O anticorrelation is unclear. HP~1 has a relatively low mass, with absolute magnitude M$_{V}$=$-$6.46,
therefore it is probably less prone to show this effect, as demonstrated
for several clusters by Carretta et al. (2009). 

The dominantly s-elements are moderately enhanced with [Ba/Fe]$\approx+0.32$
and the r-element [Eu/Fe]$\approx+0.30$.
These values are very similar to the results for NGC 6522
(Barbuy et al. 2014), indicating that these two clusters that
are located in the central parts of the Galactic bulge may
be of similar origin.


 The abundances in the earliest globular clusters
 can reveal the nature of the first
stars and also the location at which they formed in the Galaxy. It 
appears to be of great importance to further investigate this and 
other such clusters in terms of metallicity, abundances, kinematics,
 orbits, and ages.

\begin{acknowledgements}
BB, EC, AV, CSM, HE, and EB acknowledge grants and fellowships from 
CNPq, Capes and Fapesp.
SO acknowledges the financial support from the Universit\`a di Padova and
from the
Italian Ministero dell'Universit\`a e della Ricerca
Scientifica e Tecnologica (MURST), Italy. 
DM acknowledges support from the  BASAL Center for Astrophysics and Associated
 Technologies PFB-06 and FONDECYT Project 1130196.
 MZ acknowledges FONDECYT Project 1150345.
DM and MZ also acknowledge support from the Millennium Institute 
of Astrophysics MAS IC-12009.
\end{acknowledgements}


\begin{appendix}

\section{Equivalent widths and atomic data}


\begin{longtable}{cccrrrrrrrrrrrrr}
\caption{\label{EW} \ion{Fe}{I} and \ion{Fe}{II} lines, 
their wavelengths,
excitation potential (eV), oscillator strengths from NIST and VALD3 for 
\ion{Fe}{I} and Fuhr \& Wiese (2006) and Mel\'endez \& Barbuy (2009) for
\ion{Fe}{II}, and equivalent widths (m{\rm \AA}).}\\
\noalign{\smallskip}
\hline
\noalign{\smallskip} 
\hbox{\rm species} & ${\rm \lambda}$({\rm \AA}) & $\chi_{\rm ex}$(eV) 
& \hbox{log~gf$_{NIST}$} & \hbox{log~gf$_{VALD}$} & \hbox{log~gf$_{adopted}$} 
  & 2115 & 2461 & 2939 & 3514 & 5037 & 5485 &   \\ 
\noalign{\smallskip} 
\hline
\noalign{\smallskip}
      &            &         & FW06  & MB09 &  & & & & & & & \\       
\hline
FeII  &  5991.38   &  3.15   & $-$3.65 &$-$3.54 &  $-$3.54   &  ----   &  33.8   &  ----   &  24.6   &  27.5   &  26.2 \\
FeII  &  6149.25   &  3.89   & $-$2.84 &$-$2.69 & $-$2.69   &  26.3   &  32.9   &  23.5   &  31.4   &  22.3   &  33.0 \\
FeII  &  6247.56   &  3.89   &$-$2.43 &$-$2.30 &  $-$2.30   &  24.1   &  38.9   &  27.8   &  ----   &  28.0   &  46.1 \\
FeII  &  6416.93   &  3.89   &$-$2.88 &$-$2.64 &  $-$2.64   &  24.9   &  26.0   &  23.1   &  20.6   &  ----   &  33.5 \\
FeII  &  6432.68   &  2.89   &$-$3.50 &$-$3.57 &  $-$3.57   &  33.6   &  41.7   &  32.4   &  34.1   &  29.2   &  45.5 \\
FeII  &  6516.08   &  2.89   &$-$3.372 &$-$3.31 &  $-$3.31   &  45.7   &  59.9   &  47.9   &  51.7   &  44.0   &  64.0 \\
\hline
\hline
FeI   &  5525.544   &  4.231   & --- & $-$1.084 &  $-$1.084   &  ----   &  37.7   &  ----   &  ---   &  ----   &  40.5 \\
FeI   &  5543.360   &  4.218   & --- & $-$1.140 & $-$1.11     &  ----   &  45.5   &  ----   &  ---   &  62.1   &  35.2 \\
FeI   &  5554.894   &  4.549   & --- & $-$0.440 &  $-$0.440   &  ----   &  67.9   &  67.8   &  ---   &  74.4   &  48.6 \\
FeI   &  5560.211   &  4.435   & --- & $-$1.190 &  $-$1.16    &  39.4   &  29.5   &  41.3   &  ---   &  34.0   &  22.5 \\
FeI   &  5567.391   &  2.609   & --- & $-$2.564 & $-$2.67     &  75.8   &  58.1   &  77.6   &  ---   &  68.0   &  51.2 \\
FeI   &  5618.632   &  4.209   & --- & $-$1.276 &  $-$1.276   &  38.3   &  30.5   &  49.5   &  ---   &  34.1   &  34.6 \\
FeI   &  5619.595   &  4.387   & --- & $-$1.700 &  $-$1.67    &  31.1   &  ----   &  27.5   &  ---   &  23.6   &  ---- \\
FeI   &  5633.946   &  4.991   & --- & $-$0.270 &  $-$0.32    &  44.7   &  44.7   &  56.6   &  ---   &  38.4   &  33.6 \\
FeI   &  5638.262   &  4.220   & --- & $-$0.870 &  $-$0.84    &  63.4   &  62.2   &  76.1   &  ---   &  60.6   &  34.9 \\
FeI   &  5641.434   &  4.256   & --- & $-$1.180 & $-$1.15     &  58.8   &  48.3   &  59.5   &  ---   &  60.1   &  39.2 \\
FeI   &  5653.865   &  4.387   & --- & $-$1.640 & $-$1.61     &  26.2   &  ----   &  ----   &  ---   &  20.9   &  ---- \\
FeI   &  5679.023   &  4.652   & --- & $-$0.920 & $-$0.90     &  ----   &  31.8   &  34.5   &  ----   &  42.5   &  36.4 \\
FeI   &  5701.544   &  2.559   & --- & $-$2.216 & $-$2.216    &  ----   &  ----   &  ----   &  ----   &  94.0   &  70.7 \\
FeI   &  5706.096   &  4.283   & --- & $-$3.012 &  $-$3.012   &  ----   &  58.3   &  ----   &  ----   &  ----   &  ---- \\
FeI   &  5717.833   &  4.284   & --- & $-$1.130 &  $-$1.10    &  28.8   &  28.7   &  36.2   &  46.4   &  42.9   &  37.0 \\
FeI   &  6056.005   &  4.733   & --- & $-$0.460 &  $-$0.460   &  48.8   &  42.8   &  54.7   &  ---   &  50.8   &  34.6 \\
FeI   &  6078.491   &  4.796   & --- & $-$0.321 &  $-$0.321   &  59.3   &  49.5   &  52.0   &  55.9   &  56.5   &  44.3 \\
FeI   &  6079.008   &  4.652   & $-$1.100 & $-$1.120 &  $-$1.100   &  30.8   &  31.8   &  26.5   &  24.2   &  31.4   &  ---- \\
FeI   &  6082.710   &  2.223   & $-$3.573 & $-$3.573 & $-$3.573    &  67.6   &  48.2   &  61.2   &  50.9   &  53.8   &  24.5 \\
FeI   &  6093.643   &  4.608   & $-$1.470 & $-$1.599 &  $-$1.470   &  20.8   &  ----   &  ----   &  ----   &  24.2   &  ---- \\
FeI   &  6096.664   &  3.984   & $-$1.880 & $-$1.930 &  $-$1.88    &  34.4   &  23.8   &  28.9   &  23.1   &  30.5   &  ---- \\
FeI   &  6151.617   &  2.176   & $-$3.299 & $-$3.299 & $-$3.299    &  80.9   &  76.2   &  81.1   &  78.2   &  76.2   &  47.5 \\
FeI   &  6157.728   &  4.076   & $-$1.22 & $-$1.260 &  $-$1.22     &  69.6   &  60.8   &  72.4   &  40.0   &  64.0   &  41.8 \\
FeI   &  6165.360   &  4.143   & $-$1.474 & $-$1.474 &  $-$1.474   &  53.1   &  44.2   &  47.3   &  49.0   &  49.7   &  42.2 \\
FeI   &  6180.203   &  2.728   & $-$2.649 & $-$2.586 &  $-$2.649   &  84.4   &  74.9   &  87.3   &  79.7   &  79.0   &  61.7 \\
FeI   &  6187.989   &  3.943   & $-$1.67 & $-$1.720 &  $-$1.67     &  51.7   &  39.9   &  56.7   &  46.5   &  43.2   &  23.3 \\
FeI   &  6200.313   &  2.609   & $-$2.437 & $-$2.437 &  $-$2.437   &  ----   &  ----   &  ----   &  97.7   &  ----   &  ---- \\
FeI   &  6219.281   &  2.198   & $-$2.433 & $-$2.433 &  $-$2.43    &  ----   &  ----   &  ----   &  ----   &  ----   &  97.9 \\
FeI   &  6226.734   &  3.884   & --- & $-$2.220 &  $-$2.220        &  ----   &  ----   &  21.6   &  ----   &  ----   &  ---- \\
FeI   &  6229.226   &  2.845   & $-$2.805 & $-$2.805 &  $-$2.805   &  26.8   &  36.9   &  51.7   &  45.6   &  41.5   &  22.8 \\
FeI   &  6240.646   &  2.223   & $-$3.173 & $-$3.233 &  $-$3.173   &  75.7   &  54.9   &  74.6   &  ---   &  70.0   &  37.3 \\
FeI   &  6246.318   &  3.603   & $-$0.877 & $-$0.733 &  $-$0.877   &  ----   &  68.6   &  ----   &  ---   &  90.8   &  80.9 \\
FeI   &  6265.132   &  2.176   & $-$2.550 & $-$2.550 &  $-$2.550   &  ----   &  ----   &  ----   &  ----   &  ----   &  95.5 \\
FeI   &  6270.223   &  2.858   & $-$2.609 & $-$2.464 &  $-$2.609   &  76.8   &  64.5   &  68.7   &  60.7   &  66.5   &  44.1 \\
FeI   &  6271.278   &  3.332   & $-$2.703 & $-$2.703 & $-$2.703    &  24.8   &  20.8   &  24.9   &  ----   &  ----   &  ---- \\
FeI   &  6302.494   &  3.686   & --- & $-$0.973 &  $-$0.973        &  84.7   &  77.9   &  83.7   &  ---   &  85.0   &  62.2 \\
FeI   &  6311.500   &  2.832   & $-$3.141 & $-$3.141 &  $-$3.141   &  48.8   &  25.9   &  42.8   &  30.4   &  39.0   &  ---- \\
FeI   &  6315.306   &  4.143   & $-$1.232 & $-$1.232 &  $-$1.232   &  62.8   &  53.5   &  61.0   &  57.3   &  56.0   &  38.9 \\
FeI   &  6315.811   &  4.076   & $-$1.66 & $-$1.710 &  $-$1.66     &  40.0   &  ----   &  40.0   &  33.0   &  36.8   &  24.4 \\
FeI   &  6322.685   &  2.588   & $-$2.426 & $-$2.426 &  $-$2.426   &  ----   &  95.1   &  ----   &  95.8   &  ----   &  82.9 \\
FeI   &  6336.823   &  3.686   & $-$0.856 & $-$0.856 &  $-$0.856   &  ----   &  98.3   &  ----   &  ---   &  ----   &  88.8 \\
FeI   &  6344.148   &  2.433   & $-$2.923 & $-$2.923 & $-$2.92     &  ----   &  87.8   &  ----   &  88.8   &  91.0   &  62.1 \\
FeI   &  6355.028   &  2.845   & $-$2.291 & $-$2.350 &  $-$2.291   &  88.6   &  71.3   &  79.8   &  ---   &  78.3   &  51.7 \\
FeI   &  6380.743   &  4.186   & $-$1.376 & $-$1.376 & $-$1.376    &  50.0   &  45.3   &  47.4   &  33.7   &  51.1   &  31.3 \\
FeI   &  6392.538   &  2.279   & --- & $-$4.030 &  $-$4.030        &  42.5   &  21.6   &  38.0   &  24.3   &  32.6   &  ---- \\
FeI   &  6408.017   &  3.686   & $-$1.018 & $-$1.018 &  $-$1.018   &  96.7   &  86.0   &  87.8   &  84.4   &  85.4   &  77.1 \\
FeI   &  6411.648   &  3.654   & $-$0.718 & $-$0.595 &  $-$0.718   &  ----   &  ----   &  ----   &  ----   &  ----   &  98.9 \\
FeI   &  6419.949   &  4.733   & $-$0.27 & $-$0.240 &  $-$0.27     &  66.7   &  58.8   &  65.4   &  63.8   &  66.4   &  47.0 \\
FeI   &  6469.192   &  4.835   & $-$0.81 & $-$0.770 & $-$0.81      &  ----   &  26.9   &  ----   &  ----   &  38.5   &  ---- \\
FeI   &  6475.624   &  2.559   & $-$2.942 & $-$2.942 &  $-$2.942   &  76.4   &  64.5   &  81.4   &  68.6   &  83.4   &  50.4 \\
FeI   &  6481.870   &  2.279   & $-$2.984 & $-$2.984 & $-$2.984    &  96.4   &  79.4   &  98.3   &  89.2   &  ----   &  69.9 \\
FeI   &  6498.938   &  0.958   & $-$4.687 & $-$4.699 &  $-$4.687   &  ----   &  80.0   &  ----   &  94.8   &  ----   &  58.9 \\
FeI   &  6518.366   &  2.832   & $-$2.298 & $-$2.460 &  $-$2.298   &  82.0   &  68.5   &  86.8   &  ---   &  76.6   &  53.1 \\
FeI   &  6569.214   &  4.733   & $-$0.450 & $-$0.420 &  $-$0.450   &  63.4   &  60.5   &  65.2   &  67.2   &  63.1   &  46.2 \\
FeI   &  6574.226   &  0.990   & $-$5.004 & $-$5.023 &  $-$5.004   &  93.5   &  62.3   &  89.5   &  78.8   &  80.8   &  44.5 \\
FeI   &  6575.016   &  2.588   & $-$2.710 & $-$2.710 &  $-$2.710   &  95.2   &  79.8   &  93.2   &  85.7   &  83.3   &  65.3 \\
FeI   &  6581.209   &  1.485   & $-$4.679 & $-$4.679 &  $-$4.679   &  56.8   &  28.3   &  58.4   &  39.7   &  42.2   &  21.7 \\
FeI   &  6593.870   &  2.433   & $-$2.422 & $-$2.422 &  $-$2.422   &  ----   &  ----   &  ----   &  ----   &  ----   &  92.8 \\
FeI   &  6597.559   &  4.796   & $-$1.050 & $-$1.070 &  $-$1.05    &  29.5   &  20.6   &  26.4   &  31.7   &  26.3   &  ---- \\
FeI   &  6608.025   &  2.279   & --- & $-$4.030 &  $-$4.030        &  39.3   &  ----   &  38.0   &  27.9   &  35.9   &  ---- \\
FeI   &  6609.110   &  2.559   & $-$2.692 & $-$2.692 & $-$2.692    &  95.4   &  77.9   &  93.0   &  87.7   &  82.7   &  61.8 \\
FeI   &  6627.544   &  4.549   & --- & $-$1.680 &  $-$1.680        &  20.5   &  ----   &  ----   &  ----   &  ----   &  ---- \\
\noalign{\smallskip} \hline 
\end{longtable}

\end{appendix}

\end{document}